\documentclass{emulateapj}

\slugcomment{Accepted to Ap.J.}

\shortauthors{Hubeny, \& Burrows}
\shorttitle{Non-chemical equilibrium models}

\begin{document}

\title{A Systematic Study of Departures from Chemical Equilibrium 
in the Atmospheres of Substellar Mass Objects
}

\author{Ivan Hubeny\altaffilmark{1} \& Adam Burrows\altaffilmark{1}}

\altaffiltext{1}{Department of Astronomy and Steward Observatory, The University
of Arizona, Tucson, AZ 85721}

\begin{abstract}
We present a systematic study of the spectral consequences
of departures from chemical equilibrium in the atmospheres of L and T 
dwarfs, and for even cooler dwarfs.  The temperature/pressure profiles of the non-equilibrium
models are 
fully consistent with the non-equilibrium chemistry. Our grid of non-equilibrium models 
includes spectra for effective temperatures from 200 K to 1800 K, three surface gravities,  
four possible values of the coefficient of eddy diffusion in the radiative zone,
and three different CO/CH$_4$ chemical reaction prescriptions. We also provide
clear and cloudy model variants. We find, in keeping with previous studies,
that there are essentially only two spectral regions where the effects of departures
from chemical equilibrium can influence the predicted spectrum.  
These are in the M ($\sim$4$-$5 $\mu$m) and N (8$-$14 $\mu$m)
bands due to CO and NH$_3$, respectively.  The overabundance of CO
translates into flux suppressions of at most $\sim$40\% between
effective temperatures of 600 and 1800 K.  The effect is largest around
$T_{\rm eff} \approx 1100$ K.  The underabundance of ammonia 
translates into flux enhancements of no more than
$\sim$20\% for the $T_{\rm eff}$ range from 300 to 1800 K, with the largest
effects at the lowest values of $T_{\rm eff}$.  The magnitude
of the departure from chemical equilibrium increases with decreasing gravity,
with increasing eddy diffusion coefficient, and with decreasing speed of
the CO/CH$_4$ reaction.  Though these effects are modest, they lead to
better fits with the measured T dwarf spectra.  Furthermore, the suppression in 
the M band due to non-equilibrium enhancements in the CO abundance disappears
below $\sim$500 K, and is only partial above $\sim$500 K, preserving   
the M band flux as a useful diagnostic of cool atmospheres and maintaining
its importance for searches for the cooler brown dwarfs beyond the Ts.  
\end{abstract}


\section{Introduction}
\label{intro}

It has been long realized that departures from the local chemical equilibrium (LCE)
may play a significant role in the atmospheres of the substellar-mass objects
(SMO), that is of giant planets and brown dwarfs. Essentially, some chemical 
reactions occurring in these atmospheres may be very slow, so that vertical 
transport via convective motions or eddy diffusion
can lead to departures from LCE. The mechanism was first suggested to operate in
the Jovian planets of the solar system (Prinn \& Barshay 1977, Barshay \& Lewis 1978,
Fegley \& Prinn 1985, Yung et al. 1988, and Fegley \& Lodders 1994).

Fegley \& Lodders (1996) suggested that the same mechanism can also operate in the 
atmospheres of brown dwarfs. Noll, Geballe, \& Marley (1997) studied in more
detail the non-equilibrium enhancement of CO in the atmosphere of Gl 229B.
Subsequently, Griffith \& Yelle (1999) and Saumon et al. (2000) provided a more
detailed analysis. The latter paper also suggested that a non-equilibrium
depletion of NH$_3$ may be important. Saumon et al. (2003) demonstrated the 
effects of non-equilibrium chemistry for a range of effective temperatures 
($T_{\rm eff}$ between 800 and 1600 K). Recently, Saumon et al. (2006, 2007)
showed that the non-equilibrium effects in the carbon and nitrogen
chemistry have significant effects on the predicted spectrum of brown dwarfs, 
and in turn fit the observed spectrum of the T7.5 dwarf Gliese 570D (Saumon 
et al. 2006) and 2MASS 0415 (Saumon et al. 2007) much better than
models based on local chemical equilibrium.  Finally, Leggett et al. (2007)
extended the Saumon et al. (2006) models to a larger range of parameters, and
presented a comparison of model predictions with observed infrared colors.

In this paper, we extend and generalize the Saumon et al. and Leggett et al.
treatments by computing a large grid of solar-metallicity
models for effective temperatures $T_{\rm eff}$
covering the whole domain of T and L dwarfs (700 - 1800 K), as well as
for cooler dwarfs (200 $-$ 700 K), for three values of surface gravity
($\log g = 4.5,\, 5.0,\, 5.5$ cm s$^{-2}$), for several values of the coefficient
of eddy diffusion (see \S 3.1), and for three different reaction rate
timescales for the carbon reactions (see \S 3.1). We 
compute self-consistent models in the sense that the atmospheric structure
($T/P$ profile) is computed taking into account departures from chemical
equilibrium. We have computed two grids of models, one cloudless, and one
including clouds composed of silicate condensates.

In \S\ref{model}, we outline our procedures for computing model atmospheres, while
allowing for departures from chemical equilibrium. In \S\ref{results}, we present  
and analyze the predicted spectra computed with and without chemical
equilibrium for L and T dwarfs.  We explore the dependence on $T_{\rm eff}$,
gravity, and eddy mixing coefficient in the radiative zone and compare clear and 
representative cloudy models.  We also discuss the effect of non-equilibrium chemistry
for objects cooler than the coolest T dwarfs down to effective temperatures of 200 K.  
Then, in \S\ref{gl570d}, we briefly present a comparison
between theory and observation in the mid-IR for Gliese 570D. 
Finally, in \S\ref{conclusions}, we summarize our conclusions.

\section{Modeling Procedures}
\label{model}

\subsection{Non-equilibrium Carbon and Nitrogen Chemistry}

The chemistry of carbon and nitrogen in the atmospheres of SMOs is
essentially described by the net reactions
\begin{equation}
\label{co}
{\rm CO + 3H_2 \longleftrightarrow CH_4 + H_2O}\, ,
\end{equation}
for carbon, and
\begin{equation}
\label{n2}
{\rm N_2 + 3H_2 \longleftrightarrow 2 NH_3}\, ,
\end{equation}
for nitrogen.
Because of the strong C=O and N$\equiv$N bonds, the reactions (\ref{co}) and
(\ref{n2}) proceed much faster from right to left than from left
to right. For instance, for carbon the reaction
in which CO is converted to CH$_4$ is very slow, and, therefore, CO
can be vertically transported in the atmosphere to the upper and cooler
regions by convective motions or eddy diffusion. The net result is an
overabundance of CO and N$_2$ and an underabundance of NH$_3$ and CH$_4$ 
in the upper atmosphere. Since more oxygen atoms are now tied in CO,
the mixing ratio of water is also reduced because of conservation of 
the total number of oxygen atoms.

The mixing (vertical
transport) timescale, $t_{\rm mix}$, was suggested by Griffith \& Yelle (1999) 
to be parameterized by
\begin{equation}
\label{tmix1}
t_{\rm mix} = H^2/K_{zz}\, ,
\end{equation}
where $H$ is the pressure scale height, and $K_{zz}$ is the coefficient of
eddy diffusion. Equation (\ref{tmix1}) applies in the radiative (convectively
stable) zone. In the convection zone, the mixing time is given by
(Saumon et al. 2006):
\begin{equation}
t_{\rm mix} = 3 H_c/v_c\, ,
\end{equation}
where $H_c$ is the convective mixing length (adopted here to be equal to $H$),
and $v_c$ is the convective velocity.
We stress that while the mixing time is well defined in the convection zone,
it can only be roughly parameterized in the radiative zone. As discussed
e.g. by Saumon et al. (2006, 2007), the reasonable values of $K_{zz}$ are in the
range $10^2$ -- $10^6$ cm$^2$ s$^{-1}$.

The chemical reaction timescale, $t_{\rm chem}$, generally decreases rapidly 
with increasing pressure and temperature.  
In the atmospheric regions where $t_{\rm mix} \geq t_{\rm chem}$, the species
acquire their equilibrium mixing ratios. This occurs at temperatures and
pressures larger than that at the depth where $t_{\rm mix} = t_{\rm chem}$.
At smaller pressures, the species acquire mixing ratios equal to those at
the $t_{\rm mix} = t_{\rm chem}$ layer, as first suggested for the case of
the CO/CH$_4$ reaction by Prinn \& Barshay (1977). 

The non-equilibrium mixing ratios thus depend sensitively on the
chemical reaction timescales, which unfortunately are poorly known.
For the N$_2$ $\rightarrow$ NH$_3$ conversion, we adopt the timescale
given by Lodders \& Fegley (2002):
\begin{eqnarray}
t_{\rm chem}\equiv t_{\rm N_2} = \frac{1}{\kappa_{\rm N_2} N({\rm H}_2)}\, \nonumber\\
\kappa_{\rm N_2} = 8.54\times 10^{-8}\, \exp\left(-\frac{81515}{T}\right)\, ,
\end{eqnarray}
where $\kappa_{\rm N_2}$ is the rate constant for the N$_2$ $\rightarrow$ NH$_3$
reaction, and $N({\rm H}_2)$ is the number density of H$_2$.

The reaction timescales for the CO $\rightarrow$ CH$_4$ conversion are
uncertain because the correct chemical pathway is still unknown.
We will, therefore, use three different timescales, suggested earlier in
the literature:

\noindent i) the ``slow'' chemical timescale after Prinn \& Barshay (1977):
\begin{eqnarray}
t_{\rm chem}\equiv t_{\rm CO} = \frac{N({\rm CO})}{\kappa_{\rm CO} N({\rm H}_2)
\, N({\rm H_2CO})}\, \nonumber\\
\kappa_{\rm CO} = 2.3\times 10^{-10}\, \exp\left(-\frac{36200}{T}\right)\, ,
\end{eqnarray}
where $N({\rm M})$ is the number density of species M.

\noindent ii) and iii) the ``fast'' chemical timescale after Yung et al. (1988)
or Dunning et al. (1984). In both cases the timescale is given by 
\begin{equation}
t_{\rm CO} = \frac{N({\rm CO})}{\kappa_7 N({\rm H})
\, N({\rm H_2CO})}\,  ,
\end{equation}
where the rate coefficient differs in each reference. We use tabulated
values presented in Yung et al. (1988 - their Table III), and interpolate
to current temperatures. In this paper, we call the Dunning et al.
timescale ``fast1,'' and the Yung et al. timescale ``fast2.'' As we shall
show in \S 3, these two fast timescales produce very similar results.

\subsection{Atmosphere and Spectrum Modeling}

We use the updated code {\sc CoolTLUSTY}, described
in Sudarsky, Burrows, \& Hubeny (2003), Hubeny, Burrows, \& Sudarsky (2003), 
and Burrows, Sudarsky, \& Hubeny (2006 $-$ BSH), which is a variant of the
universal atmospheric code {\sc TLUSTY} (Hubeny 1988; Hubeny \& Lanz 1995). 
{\sc CoolTLUSTY} solves self-consistently a set of radiative transfer
equations for selected frequency points (typically 5000 points
logarithmically spaced between $7\times 10^{14}$ Hz and $10^{12}$ Hz),
and the equation of radiative+convective equilibrium. The atmosphere is
assumed to be in hydrostatic equilibrium. The opacities are not computed on the
fly; instead, they are interpolated from pre-calculated opacity tables for the
current values of temperature and density. Our opacity database is described
in detail by Sharp \& Burrows (2007). The atmospheric structure is
computed iteratively, applying the Newton-Raphson method
(first applied in the case of stellar atmospheres by Auer \& Mihalas 1969
with the name ``complete linearization'').
The discretized and linearized equations are re-organized in the so-called
Rybicki scheme (Mihalas 1978), which renders the computer time
linearly proportional to the number of frequencies (and not proportional
to the cube of the number of frequencies as in the case of the original complete
linearization). This allows us to avoid using the accelerated lambda iteration
(ALI) scheme to treat the radiative transfer equation, which leads to 
faster and more stable convergence, without compromising on computer time.

We have found that the most stable procedure to compute non-LCE models
is to perform several iterations (typically 6) of the linearization
method assuming chemical equilibrium (using the algorithm of
Burrows \& Sharp 1999 to determine the equilibrium abundances of all
species), and then to switch the non-LCE algorithm. 
In each subsequent iteration, one calculates the intersection point 
where the mixing time for the current $T/P$ profile equals the reaction time
(computed again for the current $T/P$ profile); the number densities
of CO, CH$_4$, H$_2$O, and NH$_3$ are set to constant values equal to
those found at the intersection points for pressures lower than the
pressure at the intersection. The opacities in these regions are modified 
accordingly. The next iteration of the linearization scheme is then performed
with the new opacities. The overall process usually converges in a few
additional iterations.

When clouds are taken into account, we employ the scheme described in
BSH. In the case of L and T dwarfs, we
represent individual cloud decks of many condenstate species by a single 
extended cloud deck of one representative condenstate. As in BSH, we take
forsterite as a representative condensate, and assume the cloud deck
to extend between the intersection of the condensation curve of
forsterite with the current $T/P$ profile, and the pressure where the local
temperature equals 2300 K (which schematically represents the condensation
curve of the most refractory condenstate), and with exponential decreases
of particle density on
both sides of the cloud. In the terminology of BSH, this is an E-type
cloud. In all the present models, we assume the modal particle size 100 $\mu$m.


\section{Results}
\label{results}

\subsection{L and T Dwarfs}
\label{res_lt}

Our effort has generated an extensive grid of
non-equilibrium model spectra for effective temperatures
from 700 K to 1800 K (in steps of 100 K), for three values of surface
gravity, $g = 10^{4.5},\, 10^{5.0}$, and $10^{5.5}$ cm s$^{-2}$, for four
values of the coefficient of eddy diffusion in the radiative zone
$K_{zz}= 10^2, 10^4, 10^6$, and $10^8$ cm$^2$ s$^{-1}$, and for three
different CO/CH$_4$ reaction time prescriptions\footnote{Electronic versions
of
these spectral models can be found online at {\tt
http://zenith.as.arizona.edu/}$\widetilde{\,\,\,}${\tt burrows}.}.
We have also provided clear and cloudy model variants. The latter assumes
an E-type forsterite cloud with a modal particle size of 100 $\mu$m.
In the present paper, we consider only solar-composition models,
where the solar elemental abundances are from Asplund, Grevesse, \&
Sauval (2006).

\begin{figure}
\plotone{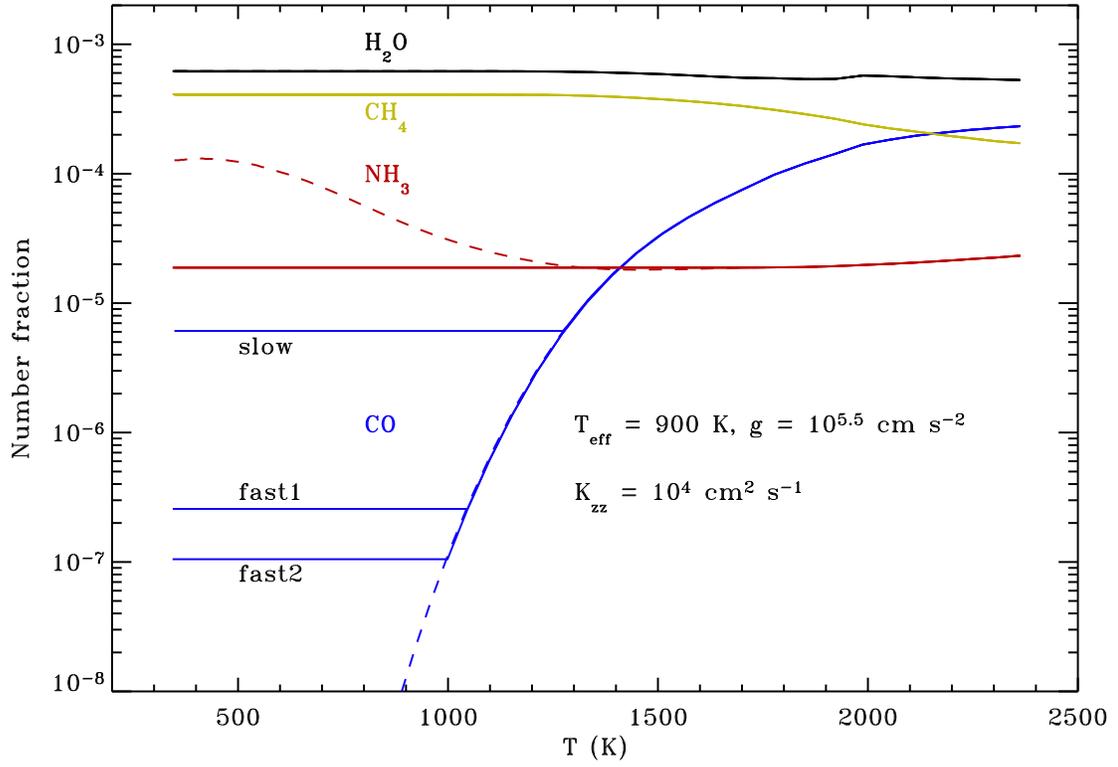}
\caption{
A comparison of the equilibrium (dashed lines) and non-equilibrium
(solid lines) abundances of water (black), methane (yellow),
ammonia (red), and CO (blue), for a representative model
with $T_{\rm eff} = 900$ K, $g=10^{5.5}$ cm s$^{-2}$, and 
$K_{zz}= 10^4$ cm$^2$ s$^{-1}$.
In this case, water and methane remain essentially at their equilibrium
abundances, while ammonia becomes underabundant at the upper layers
(leading to a lower opacity and, thus, higher predicted flux in the
8-14 $\mu$m region for non-equilibrium models). The plots also show that
CO exhibits larger departures from equilibrium with the slow reaction.
See Fig. \ref{fig5} and the text for a discussion.
}
\label{fig1}
\end{figure}

We will first discuss composition profiles and their trends with
effective temperature, surface gravity, diffusion coefficient,
and the CO/CH$_4$ reaction speed.
Figure \ref{fig1} displays the composition profiles of the
non-equilibrium species, CO, CH$_4$, NH$_3$, and H$_2$O, as a function
of the local temperature in the atmosphere, for a representative model
with $T_{\rm eff} = 900$ K, $g=10^{5.5}$ cm s$^{-2}$, and $K_{zz}= 10^4$
cm$^2$ s$^{-1}$.  The full lines show the non-equilibrium mole fractions or,
equivalently, the number fractions (expressed as a fraction of the total 
number of particles), while the dashed lines show the equilibrium
number fractions. For this model, the non-equilibrium effects in 
water and methane are negligible. The non-equilibrium number fraction of ammonia
is smaller than the equilibrium one for $T<1200$ K, as first shown by 
Saumon et al. (2006). The non-equilibrium mixing ratio of CO is significantly
larger than the equilibrium one, which in turn leads to a significant
strengthening of the 4.7-$\mu$m feature (and, to a lesser extent, to other
CO features at 3.4 and 2.8 $\mu$m) in non-equilibrium models.
Moreover, the non-equilibrium mixing ratio of CO is larger for the slower
CO/CH$_4$ reaction, because the intersection of the $t_{\rm mix}(T)$
curve with the $t_{\rm CO}(T)$ curve occurs deeper in the atmosphere
for higher temperature, at which the equilibrium CO mixing ratio is higher
(see also Fig. \ref{fig5}). Figure \ref{fig1} is analogous to Fig. 3 of
Saumon et al. (2006).

\begin{figure}
\plotone{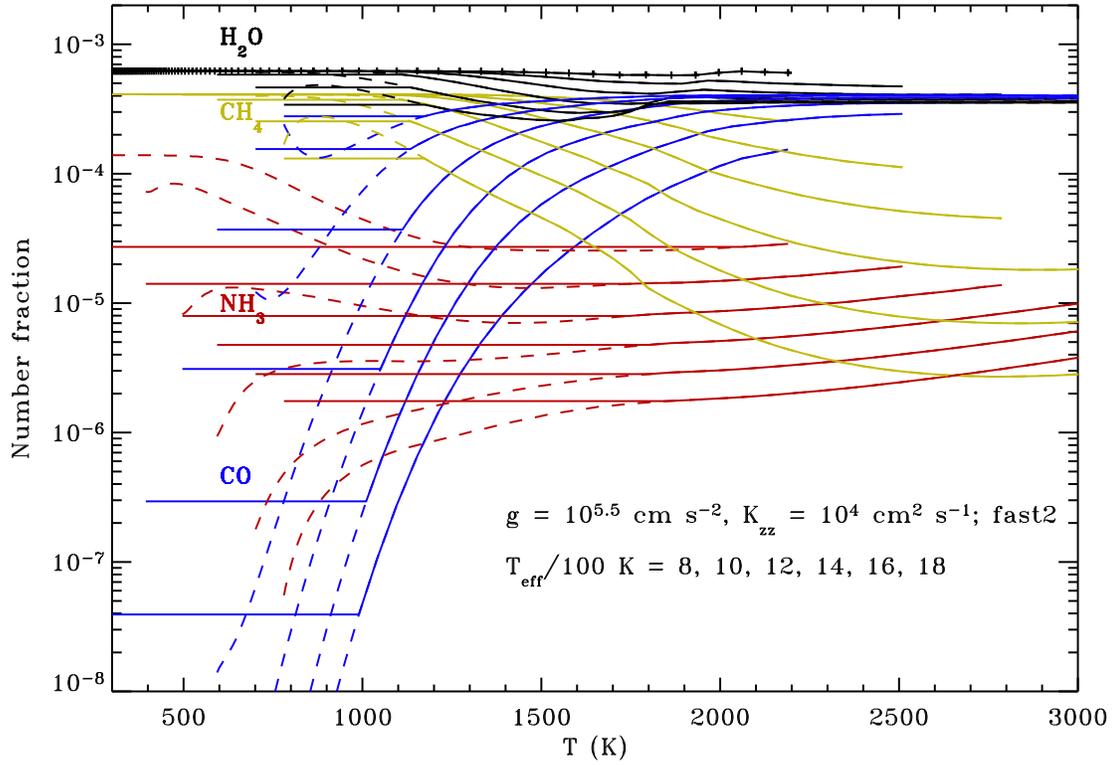}
\caption{
Similar to Fig. \ref{fig1}, but with a range of effective temperatures to display
the trends with $T_{\rm eff}$. The color scheme is
the same as in Fig. \ref{fig1}. The association of the individual lines with
effective temperatures is made by noting that in the plot the lines corresponding 
to progressively lower effective temperatures end at progressively lower
local temperatures. (In other words, the longer the lines go to
the right, the higher the effective temperature.) In all cases displayed
here, water and methane remain at their equilibrium abundances.
The degree of departure from equilibrium for ammonia decreases with increasing
effective temperature. 
}
\label{fig2}
\end{figure}

To depict the trend of composition profiles with effective temperature, we
display in Fig. \ref{fig2} composition profiles as in Fig. \ref{fig1}, but
for $T_{\rm eff}/100\, {\rm K} = 8,10,12,14,16$, and 18. To avoid cluttering 
the figure, the models are not labeled by the values of $T_{\rm eff}$. 
Instead, the models are distinguished by the fact that 
the higher the effective temperature is, the longer the curves 
continue toward high temperatures (i.e., to the right side of the plot).
First, we stress that the equilibrium abundance patterns follow from
the basic chemistry (Burrows \& Sharp 1999; Lodders \& Fegley 2002),
and are well understood.
Briefly, the same local temperature is reached at lower pressure for
models with higher effective temperatures (see Fig. \ref{fig7} for an
explicit demonstration). Due to Le Chatelier's principle,  
at a given local temperature the abundance of CH$_4$ 
decreases and the abundance of CO increases with increasing $T_{\rm eff}$. 
Analogously, the abundance of NH$_3$ decreases, and the abundance of N$_2$
(not shown here) increases with increasing $T_{\rm eff}$. These effects
are clearly shown in Fig. \ref{fig2}.

For all models displayed here, the non-equilibrium effects in water 
and methane are negligible. The non-equilibrium abundance of CO is 
always larger than the equilibrium abundance for 
$T \stackrel{\sim}{<} 1000$ K. For very low $T_{\rm eff}$, although 
the non-equilibrium abundance of CO is significantly larger than 
the equilibrium abundance, it is still small in absolute value, and, 
therefore, does not have much of an effect on the model structure 
and predicted spectra. For the highest $T_{\rm eff}$, the equilibrium
abundance profile of CO becomes flatter, and, consequently,
the differences between the non-equilibrium and equilibrium abundance 
profiles of CO decrease, and so do the differences in the predicted 
spectra. This is demonstrated in Fig. \ref{fig8}. For ammonia, departures 
from equilibrium become smaller for higher effective temperatures, because 
the equilibrium ammonia composition profile becomes flat. Therefore, fixing 
the ammonia number fraction at the intersection of the $t_{\rm mix}$ 
and the $t_{\rm NH_3}$ curves does not lead to a significant difference 
in the composition. Therefore, the non-equilibrium effects in ammonia, and, 
consequently, in the predicted flux in the 8 - 14 $\mu$m region, decrease 
with increasing $T_{\rm eff}$. This also is demonstrated in Fig. \ref{fig8}. 
Moreover, because of its diminished number fraction, ammonia becomes
a less important source of opacity as the temperature increases.

\begin{figure}
\plotone{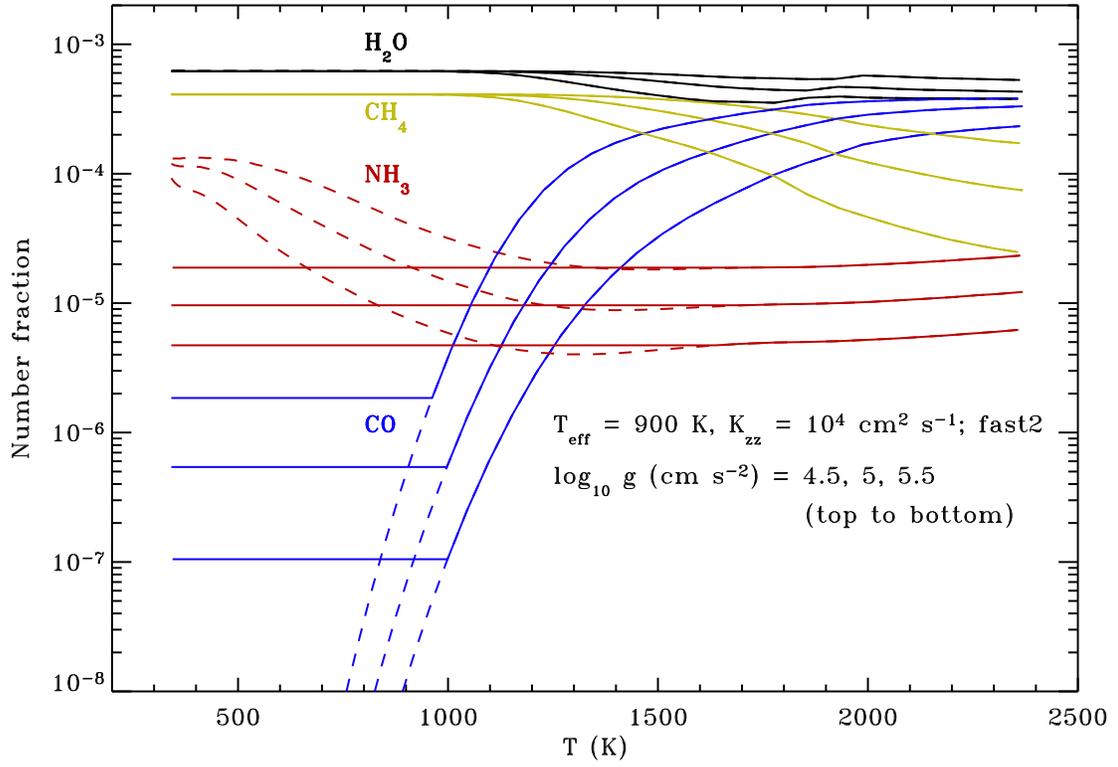}
\caption{
Similar to Fig. \ref{fig2}, but showing the trend with changing surface gravity.
This figure demonstrates that the non-equilibrium abundance of CO increases 
with decreasing gravity. See explanation in the text.
}
\label{fig3}
\end{figure}

\begin{figure}
\plotone{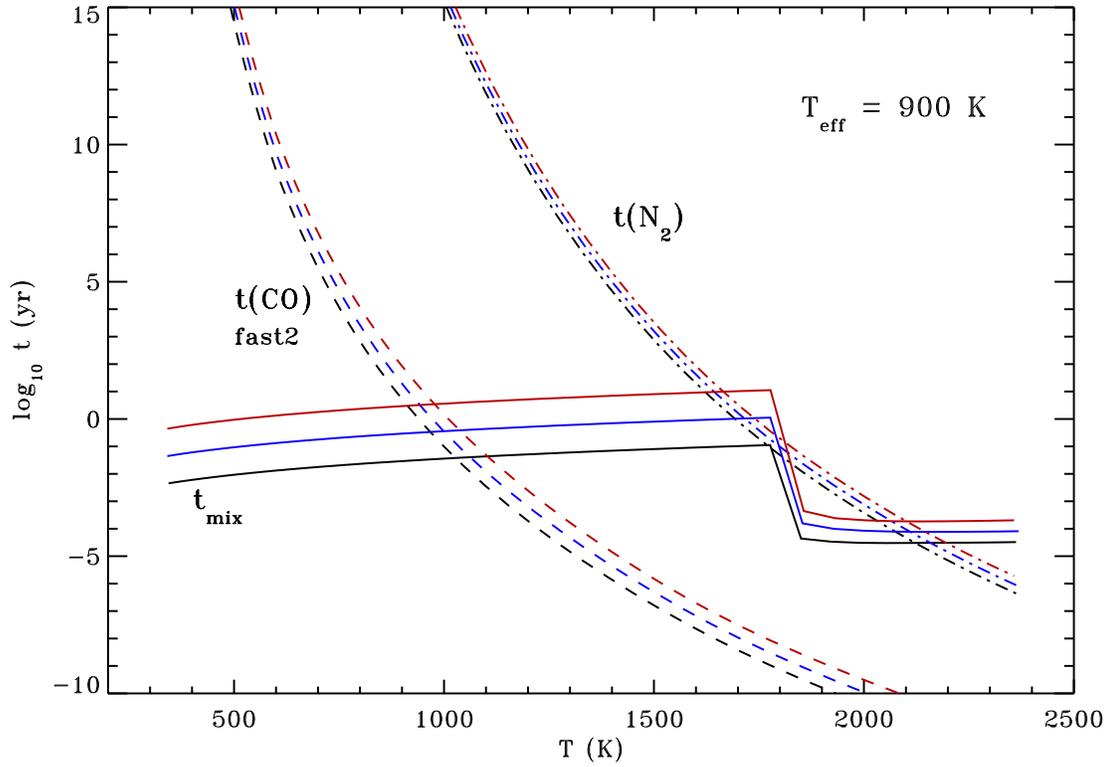}
\caption{
A plot of the mixing time (solid lines), the reaction time
for the CO/CH$_4$ (dashed lines), and the reaction time for
the NH$_3$/N$_2$ reaction (dash-dotted lines), for models with
$T_{\rm eff} = 900$~K, $K_{zz}=10^4$ cm$^2$ s$^{-1}$, and $\log_{10} g$ (cm s$^{-2}$)
= 5.5 (black), 5.0 (blue), and 4.5 (red). The abrupt change of the 
mixing-time curves represents the boundary of the convection zone.
}
\label{fig4}
\end{figure}

\begin{figure}
\plotone{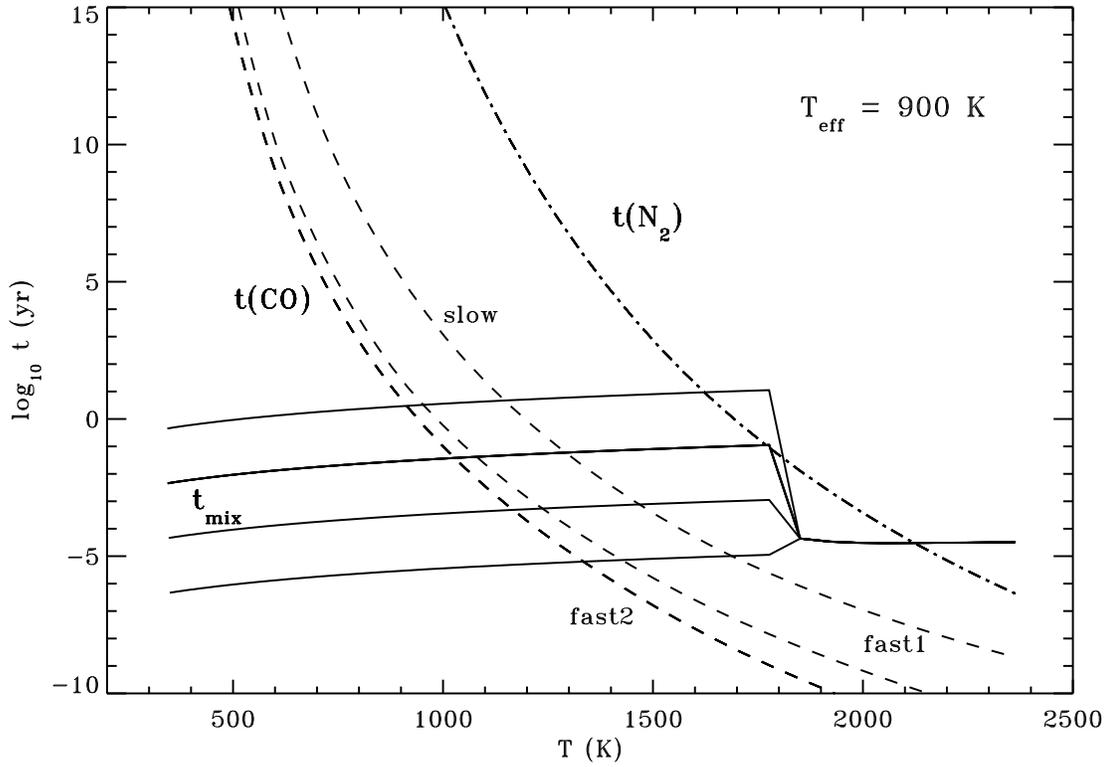}
\caption{
Similar to Fig. \ref{fig4}, but depicting
the mixing times (solid lines) corresponding to (from top to bottom)
the four values of the diffusion coefficient, 
$K_{zz}=10^2, \, 10^4,\, 10^6,\, 10^8$ cm$^2$ s$^{-1}$,
for models with $T_{\rm eff}=900$ K, and $g=10^{5.5}$ cm s$^{-2}$.
Also shown are the reaction times
for CO/CH$_4$ (dashed lines) chemistry, for the three different treatments
of the reaction speed, ``slow,'' ``fast1,'' and ``fast2'' 
(the curves are labeled accordingly), and the reaction 
times for the NH$_3$/N$_2$ reaction (dot-dashed lines).
}
\label{fig5}
\end{figure}

In Fig. \ref{fig3}, we show the trend of the composition
profiles with the surface gravity, for $T_{\rm eff} = 900$~K, 
$K_{zz} = 10^4$ cm$^2$ s$^{-1}$, and the fast CO/CH$_4$ reaction. 
The intersection of the 
$t_{\rm mix}$ and the $t_{\rm CO}$ curves occurs at essentially the same
temperature for all three surface gravities (see Fig. \ref{fig4}).
However, at a given temperature, the model with the lowest gravity exhibits
the lowest pressure, and, therefore, the CO mixing ratio at the intersection is
higher for the lower-gravity models. Consequently, the effects of departures
from chemical equilibrium are greater for the low-gravity models.

Figure \ref{fig4} displays the mixing time and the reaction times for the
models displayed in Fig.~\ref{fig3}. Notice the abrupt change of the mixing
time at $T\approx 1800$ K; this corresponds to the onset of convection,
where the mixing times are about five orders of magnitude faster.
To explain the behavior of models with different coefficients of eddy
diffusion, and with different reaction timescales for the CO/CH$_4$ reaction,
we display in Fig. \ref{fig5} the mixing times for four different diffusion
coefficients, $K_{zz} = 10^2, 10^4, 10^6$, and $10^8$ cm$^2$ s$^{-1}$.
By increasing the coefficient of eddy diffusion, $K_{zz}$, the mixing time 
decreases and, consequently, the intersection of the chemical reaction timescale
curves and the mixing time curve occurs at higher temperatures. Therefore,
the non-equilibrium mixing ratio of CO is higher. For the N$_2$/NH$_3$ reaction,
and for $K_{zz} > 10^4$ cm$^2$ s$^{-1}$, the chemical time curve 
intersects the mixing time curve in the convection zone at very high
temperature ($T \approx 2100$ K).

\begin{figure}
\plotone{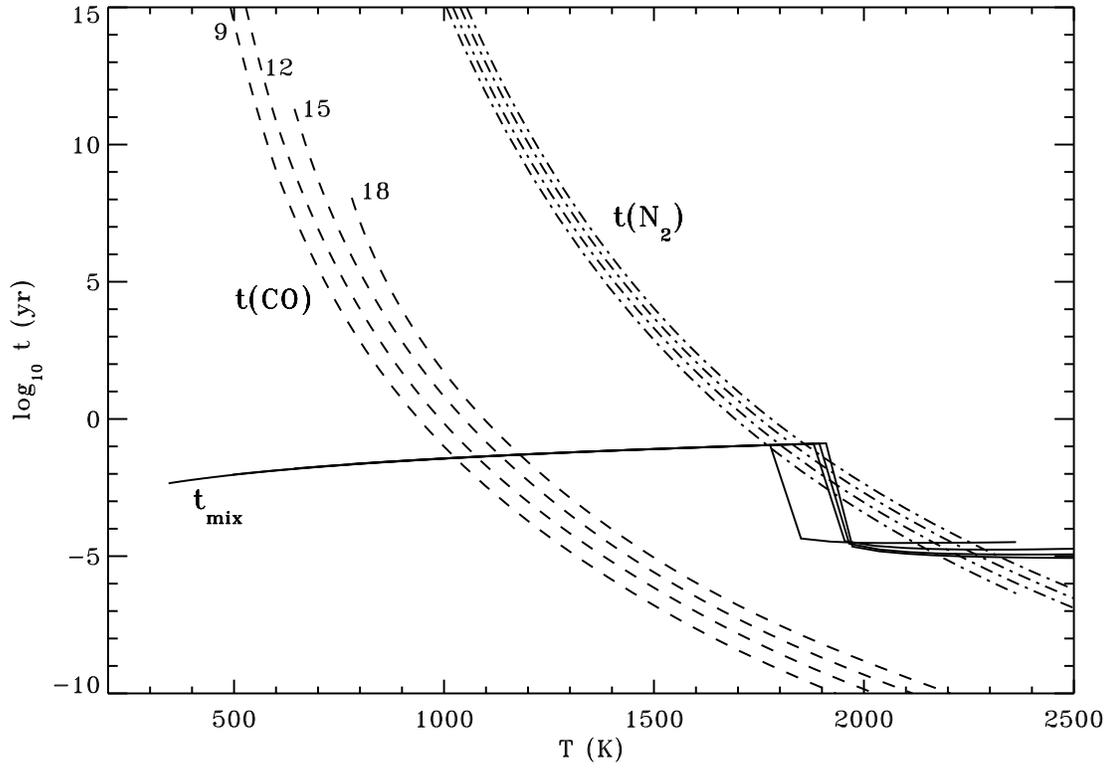}
\caption{
Similar to Fig. \ref{fig4}, but depicting
the mixing times (solid lines), and the reaction times
for CO/CH$_4$ (dashed lines) for the ``fast2'' reaction rate, and the reaction 
time for the NH$_3$/N$_2$ reaction (dot-dashed line),
for a set of models with 
$g=10^{5.5}$ cm s$^{-2}$,  $K_{zz}=10^4$ cm$^2$ s$^{-1}$, and
$T_{\rm eff}/100$ K = 9, 12, 15, 18. The reaction times for
CO/CH$_4$ are labeled accordingly.
}
\label{fig6}
\end{figure}

\begin{figure}
\plotone{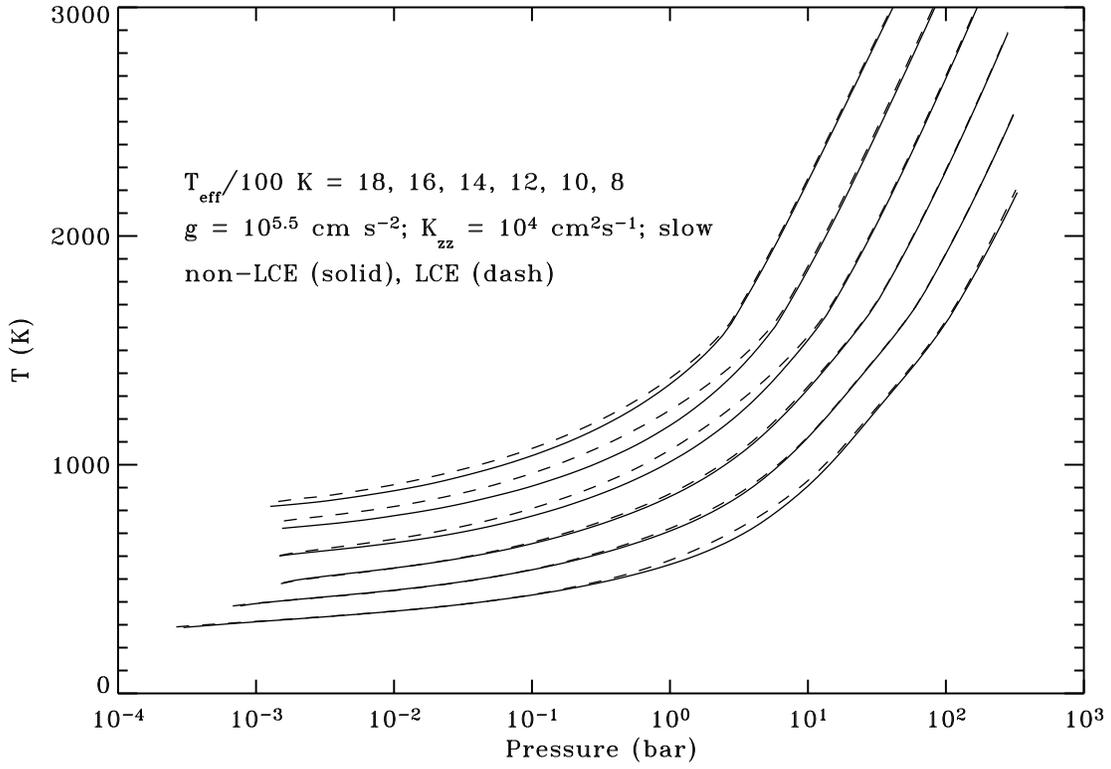}
\caption{
A comparison of temperature-pressure profiles for equilibrium (dashed lines)
and non-equilibrium (solid lines) models of $T_{\rm eff}=800,\, 1000,\, 1200,\, 1400, \, 1600$, 
and 1800 K (from bottom to top). $g$ is equal to $10^{5.5}$ cm s$^{-2}$.
}
\label{fig7}
\end{figure}

\begin{figure}
\includegraphics[width=15cm,height=17cm,keepaspectratio=false]{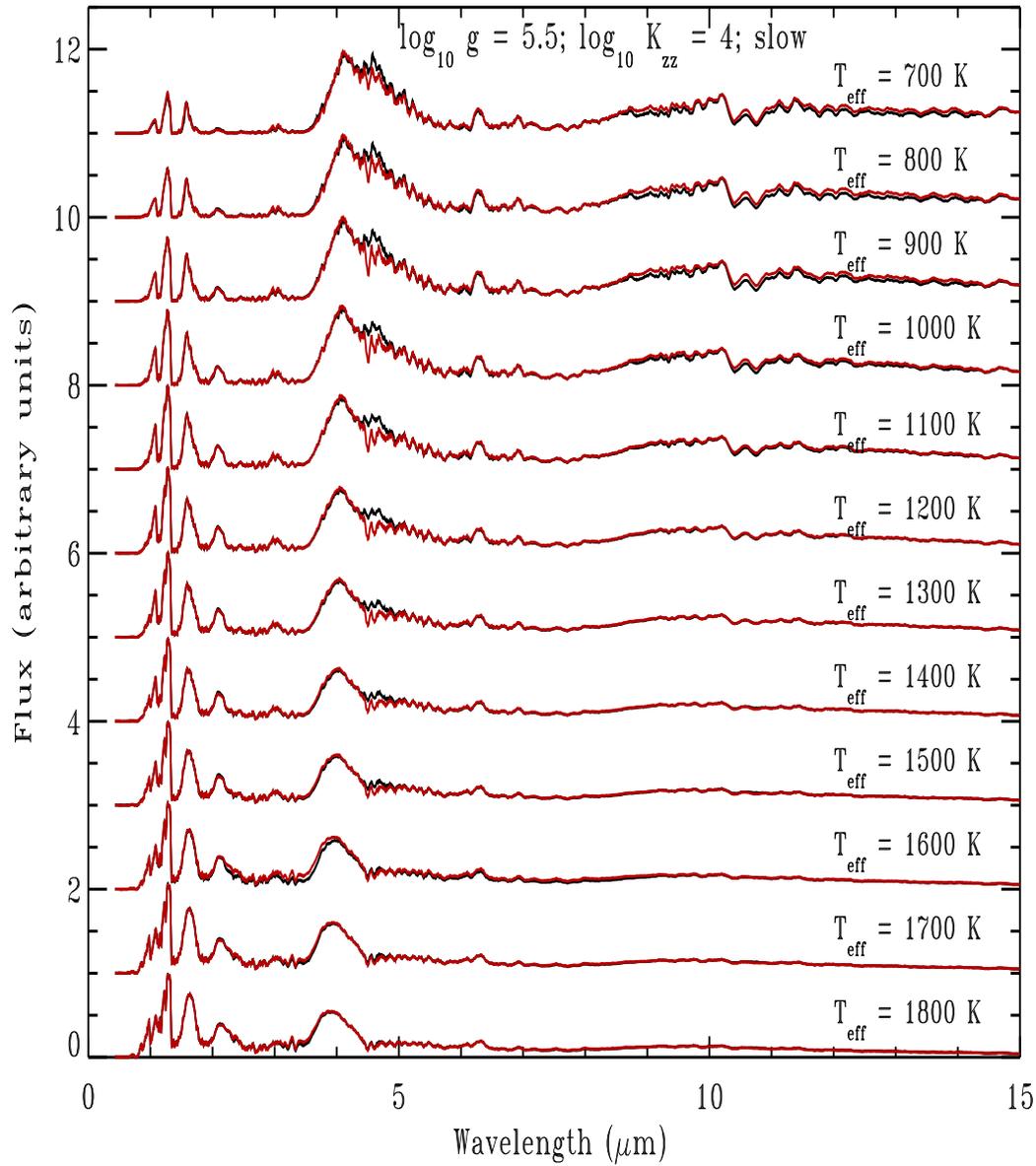}
\caption{
Synoptic comparison of model atmosphere spectra computed with 
chemical equilibrium (black lines), and with departures from 
equilibrium (red lines), for a representative value of $g = 10^{5.5}$ cm s$^{-2}$,
and $K_{zz} = 10^4$ cm$^2$ s$^{-1}$,
for a ``slow'' CO/CH$_4$ reaction, and for effective
temperatures between 700 and 1800 K, covering most of the T and 
L spectral types. There are essentially two wavelength regions where
the departure from chemical equilibrium influences the predicted spectrum:
the region around 4.7 $\mu$m (due to non-equilibrium abundances of CO), 
and a wide region between 8 and 14 $\mu$m, due to non-equilibrium
abundances of NH$_3$. The CO feature has its peak for effective
temperatures between 800 K and 1200 K; the effect generally decreases
for hotter models and essentially disappears at 1800 K.
The figure is similar to that presented by Saumon et al. (2003).
}
\label{fig8}
\end{figure}

In Fig. \ref{fig6}, we display trends of the chemical reaction time with
the effective temperature. The N$_2$/NH$_3$ reaction timescale is relatively
insensitive to the model effective temperature, while for the CO/CH$_4$ reaction,
the intersection of the curves moves to higher temperatures and pressures for
higher effective temperatures. It would seem at first sight that the departures
from chemical equilibrium would increase with increasing $T_{\rm eff}$.
However, as we have shown in Fig. \ref{fig2}, the abundance profile of CO becomes 
flatter and, thus, the departures from chemical equilibrium decrease instead.

In Fig. \ref{fig7} we display the temperature/pressure profiles
for selected models of our grid. The differences in the local temperature between
the equilibrium and non-equilibrium models are not large, but nevertheless
appreciable.

\begin{figure}
\includegraphics[width=15cm,height=17cm,keepaspectratio=false]{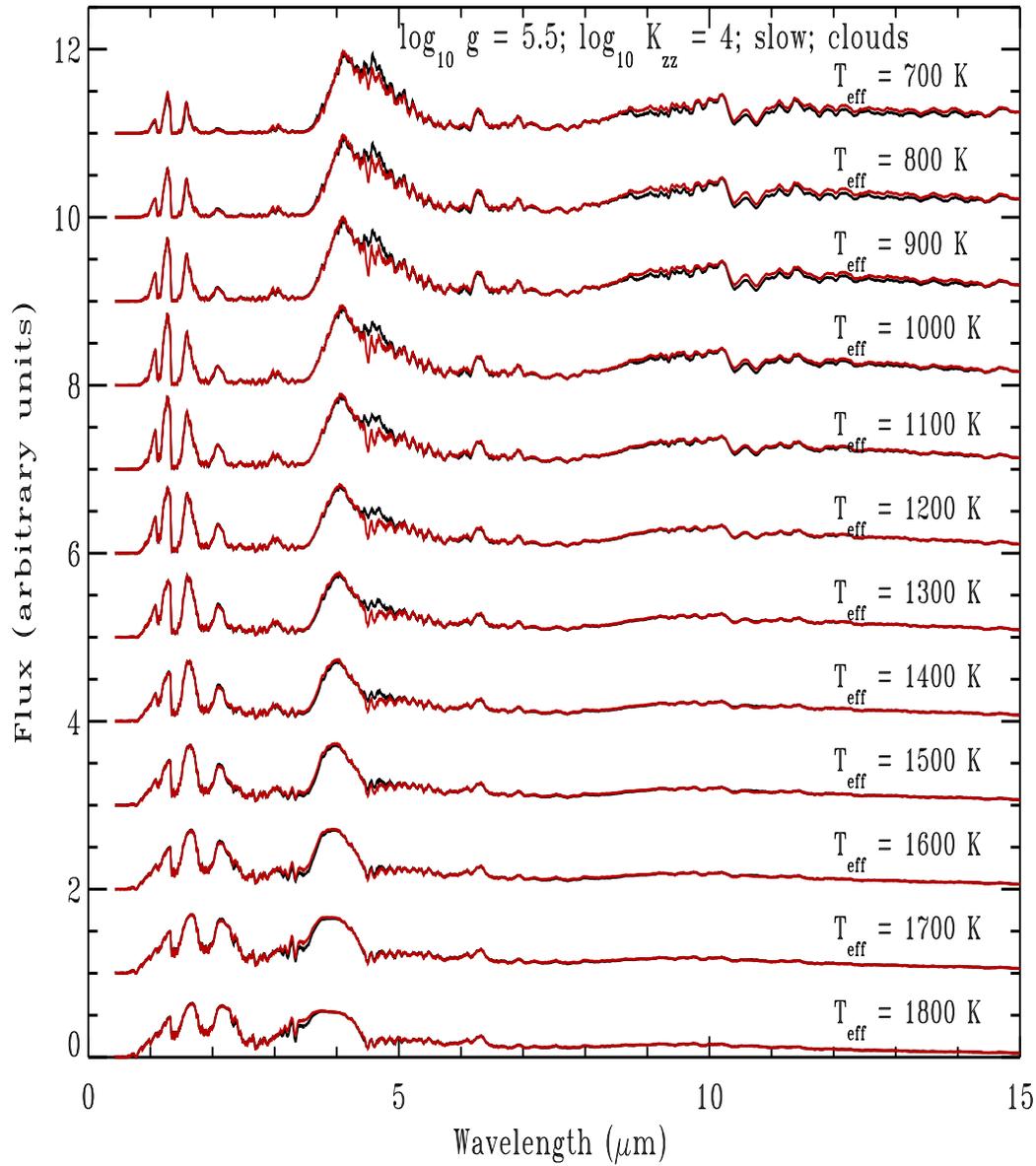}
\caption{
The same as Fig. \ref{fig8}; the only difference is that the models are computed
with clouds. We assume a cloud composed of a collection of magnesium 
silicate componds represented by a single species, in this case
forsterite, with an extension of the cloud layer toward
higher pressures, as suggested by BSH (E-type cloud model in
their notation). We assume a modal particle size of 100 $\mu$m.
The presence of clouds leads to differences in the predicted spectrum,
in particular for wavelengths below 2.5 $\mu$m, but otherwise the
effect of departures from chemical equilibrium is essentially 
the same as for the cloudless models.
}
\label{fig9}
\end{figure}

We now turn to emergent spectra.
In Fig. \ref{fig8}, we display the emergent spectra for all effective temperatures,
and for one representative value of $g = 10^{5.5}$ cm s$^{-2}$, for $K_{zz}=10^4$
cm$^2$ s$^{-1}$, and for the ``slow'' reaction rate of the CO/CH$_4$ reaction
of Prinn \& Barshay (1977). 
There are essentially only two wavelength regions where
the departures from chemical equilibrium influence the predicted spectrum,
the region around 4.7 microns (due to non-equilibrium abundances of CO), 
and a wide region between 8 and 14 microns, due to non-equilibrium
abundances of NH$_3$. The CO feature has its peak for effective
temperatures between 800 K and 1200 K; the effect generally decreases
for hotter models and essentially disappears at 1800 K. 
The non-equilibrium effect of ammonia already disappears for 
$T_{\rm eff} \stackrel{>}{\sim} 1200$ K. The behavior of the associated spectra
is readily understood upon examination of the composition 
profiles displayed in Figs. \ref{fig1} - \ref{fig3} (see also the related
discussion).
Figure \ref{fig8} is a synoptic display of the overall model trends,
but is not meant to communicate details of the predicted spectra. Therefore,
we later present in Figs. \ref{fig11}, \ref{fig12}, and \ref{fig13} enlarged plots
of the two non-equilibrium regions, the CO region between 3 and 5 $\mu$m,
and the ammonia region between 8 and 14 $\mu$m.
In these and subsequent figures that display the
predicted spectra, we smooth the the computed spectrum by a 10-point boxcar
average, which corresponds to a resolution of about 1/100.
The corresponding models with clouds are displayed in Fig. \ref{fig9}
and behave analogously, thus demonstrating that the presence of clouds
does not lead to any qualitative differences in the magnitude of departures
from chemical equilibrium. 

\begin{figure}
\plotone{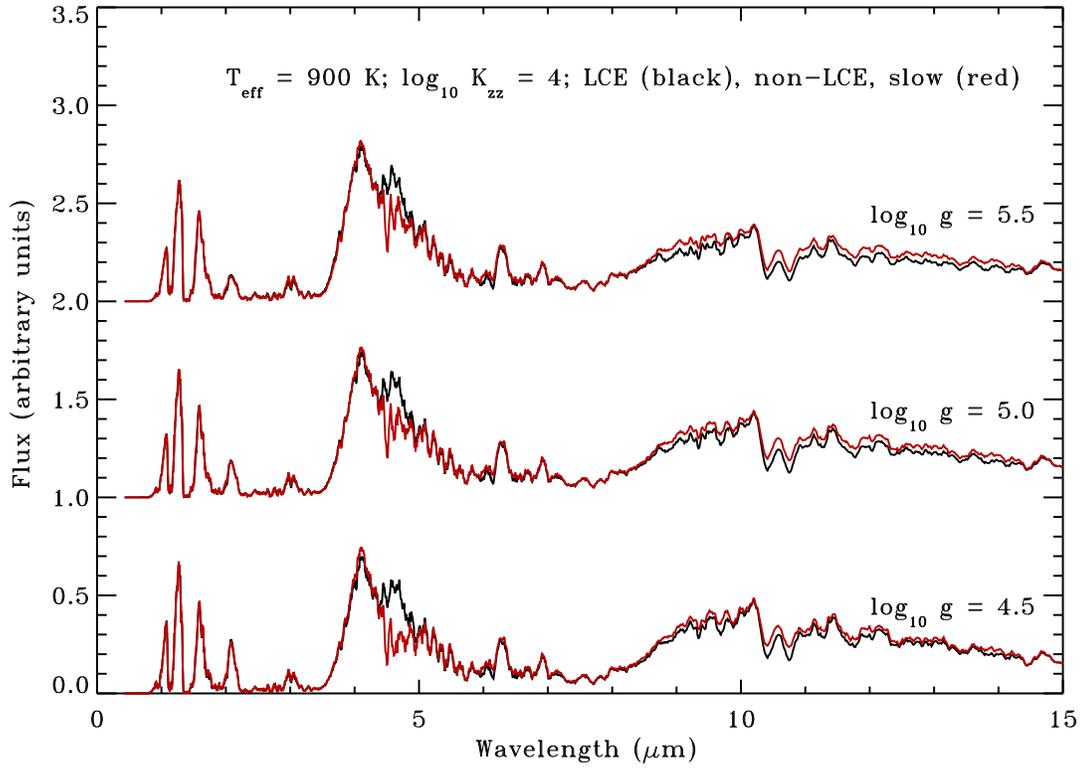}
\caption{
A comparison of predicted spectra for a model with $T_{\rm eff}=900$ K,
$K_{zz} = 10^4$ cm$^2$ s$^{-1}$, and three different surface gravities. 
We display the equilibrium models (black lines), and non-equilibrium models
(red lines) assuming the ``slow'' CO/CH$_4$ reaction rate of Prinn \&
Barshay (1977).
The plot also clearly shows that the effects of departures from
chemical equilibrium are more pronounced for lower gravities.
See Fig. \ref{fig1} and the text for explanations.
}
\label{fig10}
\end{figure}

\begin{figure}
\plotone{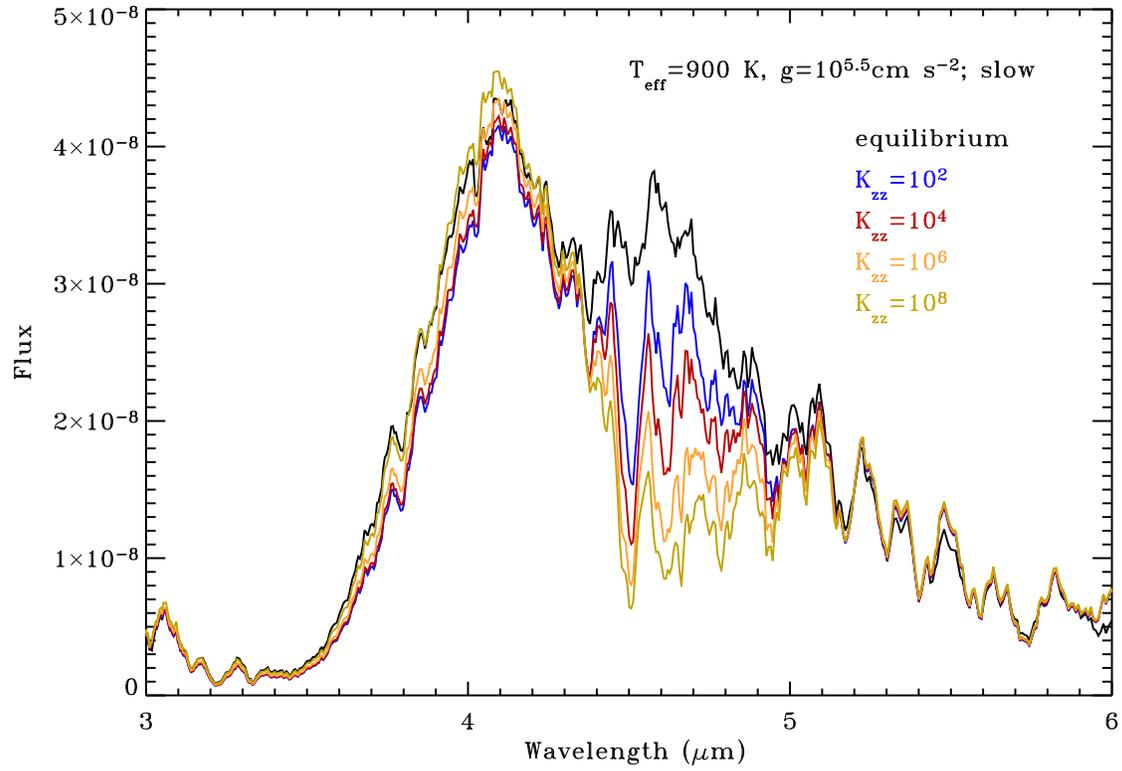}
\caption{A more detailed view of the CO non-equilibrium
region between 3 and 5 $\mu$m, for models with $T_{\rm eff}=900$ K
and $g=10^{5.5}$ cm s$^{-2}$. Displayed are the equilibrium model (black), as well
four non-equilibrium models with $K_{zz}= 10^2$ (blue), $10^4$ (red), $10^6$
(yellow), and $10^8$ (green) cm$^2$ s$^{-1}$. 
All models are for the ``slow'' reaction rate for the CO/CH$_4$ reaction 
of Prinn \& Barshay (1977). 
}
\label{fig11}
\end{figure}

\begin{figure}
\plotone{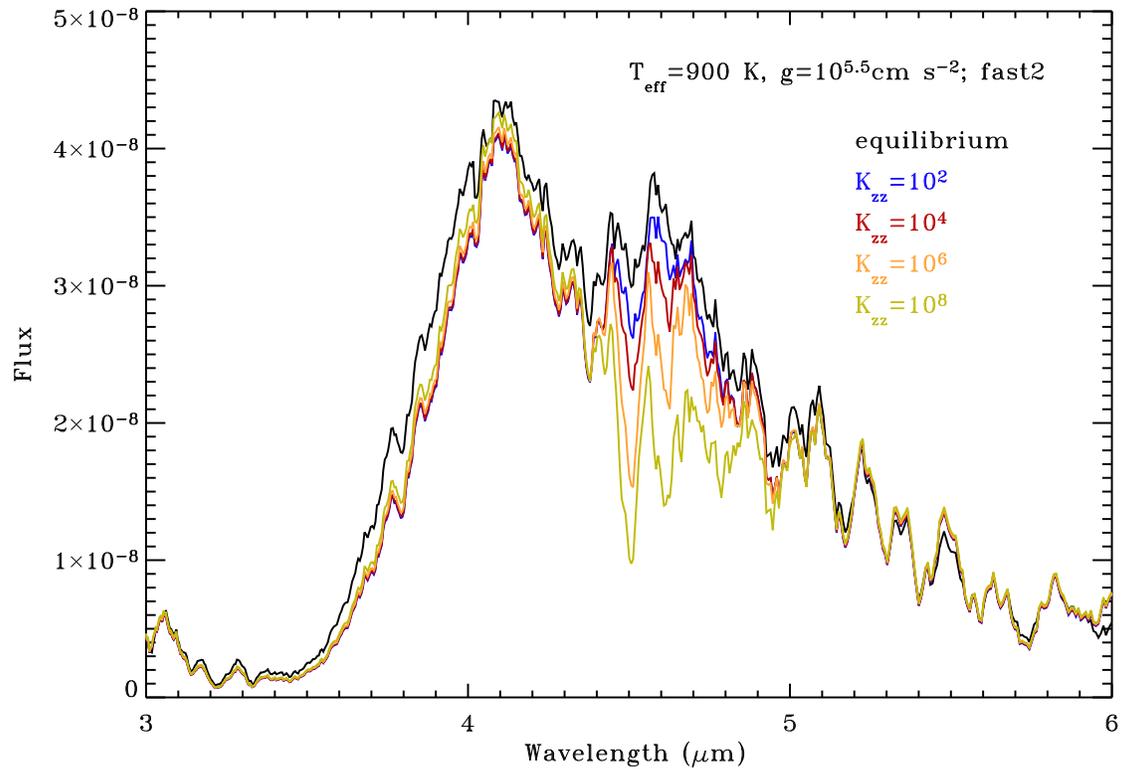}
\caption{Similar to Fig. \ref{fig11}, but for the ``fast2'' CO/CH$_4$ 
reaction rate of Yung et al. (1988). Departures from equilibrium are now
smaller, as explained in the text. 
}
\label{fig12}
\end{figure}

\begin{figure}
\plotone{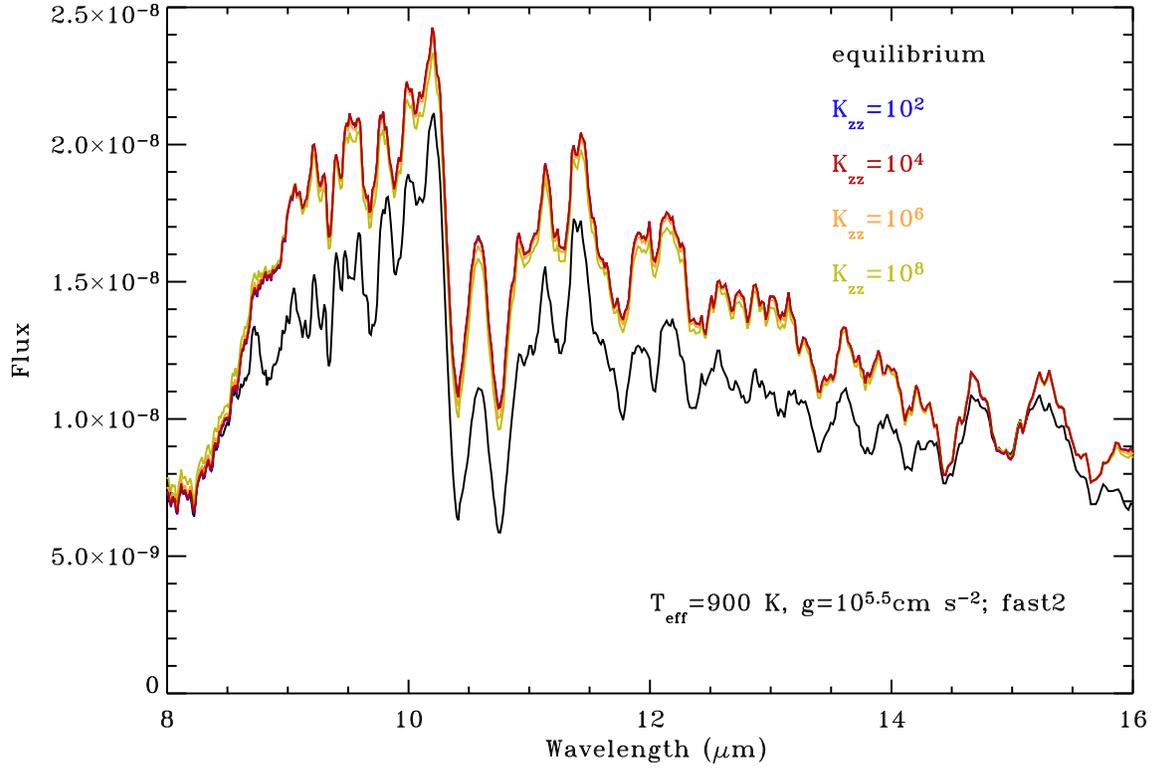}
\caption{A more detailed view of the
ammonia region between 8 and 14 $\mu$m, for the same models as displayed in
Fig. \ref{fig12}. The curves for all non-equilibrium models
essentially coincide, except for the highest value of $K_{zz}=10^8$
cm$^2$ s$^{-1}$.
The flux in the region between 8 and 14 $\mu$m is relatively
insensitive to the value of $K_{zz}$.
}
\label{fig13}
\end{figure}

Figure \ref{fig10} shows the sensitivity of the predicted spectrum to
the surface gravity. The non-equilibrium 
effects generally increase with decreasing gravity, as was explained in
the discussion of Fig. \ref{fig3}. 
Figure \ref{fig11} presents a more detailed view of the CO non-equilibrium
region between 3 and 5 $\mu$m, for one particular model with $T_{\rm eff}=900$ K
and $g=10^{5.5}$ g cm$^{-2}$. We display both the LCE model and
four non-equilibrium models with $K_{zz}= 10^2, \, 10^4, \, 10^6$, and
$10^8$ cm$^2$ s$^{-1}$. All models are for the ``slow'' reaction rate
of Prinn \& Barshay (1977) for the CO/CH$_4$ reaction. An analogous plot
for the models with the ``fast2'' reaction of Yung et al. (1988) is displayed
in Fig.~\ref{fig12}. The magnitude of departures from LCE increases significantly
with increasing $K_{zz}$. This is because the mixing time decreases with
increasing $K_{zz}$ and, thus, the intersection of the $t_{\rm mix}$ and the
$t_{\rm CO}$ curves occurs deeper in the atmosphere, where the mixing ratio
of CO is larger. This was already demonstrated in Fig. \ref{fig5}.
As Fig. \ref{fig12} shows, the departures from LCE are smaller for the
faster CO/CH$_4$ reaction rate, as demonstrated previously in Fig. \ref{fig1}.

In Fig. \ref{fig13}, we display a more detailed view of the
ammonia region between 8 and 14 $\mu$m, for the same models as displayed in
Fig. \ref{fig12}. The departures from chemical equilibrium are essentially
the same as predicted by Saumon et al. (2006, 2007).
Unlike the case of CO, the abundance of ammonia and,
therefore, the flux in the region between 8 and 14 $\mu$m, is relatively
insensitive to the value of $K_{zz}$. This is because the abundance profile
of ammonia is rather flat (see Figs. \ref{fig1} - \ref{fig3}), and, therefore,
the depth of intersection of the $t_{\rm N_2}$ and the $t_{\rm mix}$ curves
does not matter very much. The departures of 
ammonia from LCE are largely insensitive to the speed of the CO/CH$_4$
reaction, as expected. We do not show this in the figure because
the corresponding plots are essentially identical. 
The predicted departures of NH$_3$ from LCE seem to be quite
robust. 

\begin{figure}
\plotone{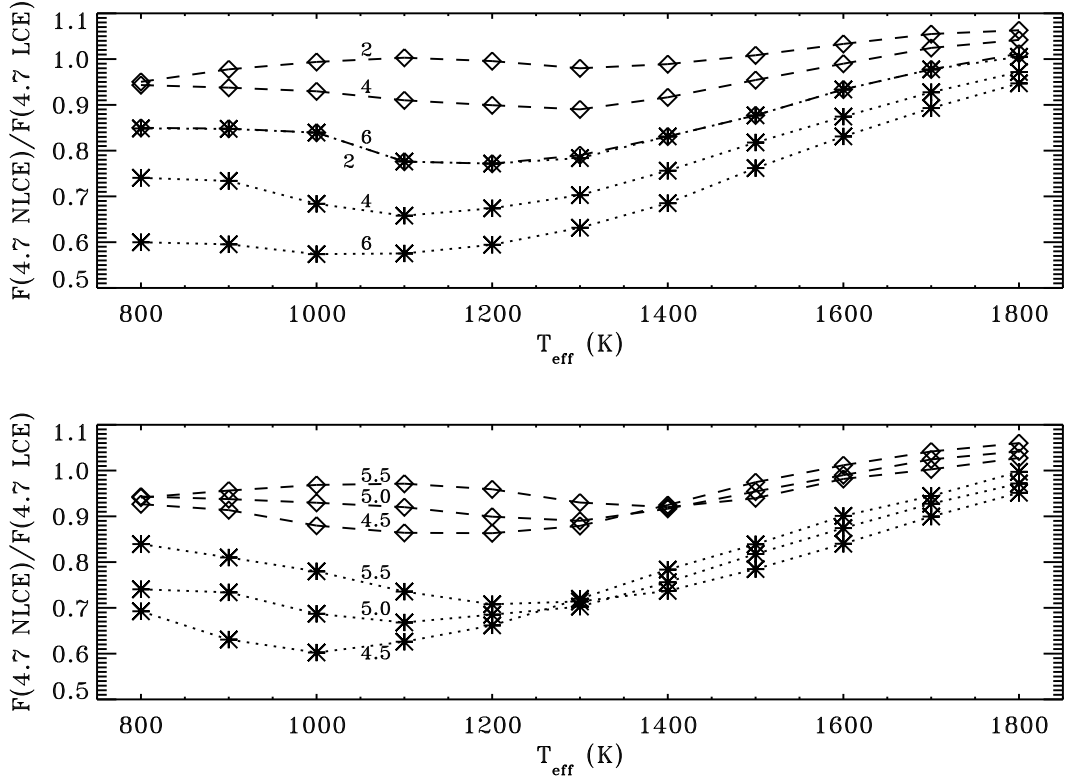}
\caption{Departures from chemical equilibrium in the 4.7-$\mu$m CO region.
The plot displays the ratio of the integrated flux in the region between
4.4 $\mu$m and 5 $\mu$m of the non-equilibrium and equilibrium models, 
as a function of effective temperature. The upper panel displays the
dependence of the non-equilibrium ratio for $g=10^5$ cm s$^{-2}$, for
the ``slow'' CO/CH$_4$ reaction (the upper set of diamonds connected
by dashed lines) and the ``fast2'' CO/CH$_4$ reaction (the lower set
of stars connected by dotted lines); each set is shown for three values of
$K_{zz}$. The curves are labelled by the value of $\log_{10} K_{zz}$
in cm$^2$ s$^{-1}$. The curves for $K_{zz}=10^6$ cm$^2$ s$^{-1}$ for
the ``fast2'' CO/CH$_4$ reaction and for $K_{zz}=10^2$ cm$^2$ s$^{-1}$ for
the ``slow'' CO/CH$_4$ reaction essentially coincide.
The lower panel displays the dependence of the non-equilibrium ratio
on surface gravity for two sets of models. In all models, $K_{zz}=10^4$
cm$^2$ s$^{-1}$. The upper set of models (diamonds connected by dashed
lines) are for the ``fast2'' CO/CH$_4$ reaction, while the lower set
of models (stars connected by dotted lines) are for the ``slow''
CO/CH$_4$ reaction. The curves are labelled by the value of $\log_{10} g$  
in cm s$^{-2}$.
}
\label{fig14}
\end{figure}

To better display departures from chemical equilibrium, we plot
in Fig. \ref{fig14} the ratio of the integrated flux between
4.4 $\mu$m and 5 $\mu$m of the non-equilibrium and equilibrium models, 
as a function of effective temperature. The upper panel displays the
dependence of the non-equilibrium ratio for $g=10^5$ cm s$^{-2}$, for
the ``slow'' CO/CH$_4$ reaction, and the ``fast2'' CO/CH$_4$ reaction. 
Each set of models is computed for three values of $K_{zz}$. The curves 
are labelled by the value of $\log_{10} K_{zz}$ in cm$^2$ s$^{-1}$.
As discussed above, assuming the ``slow'' CO/CH$_4$ reaction increases
the departures from LCE significantly, reaching about 30\% at 
$T_{\rm eff}=1100$ K (at $K_{zz}=10^4$ cm$^2$ s$^{-1}$), 
and decreasing toward higher temperatures.
For the ``slow2'' reaction, the maximum effect is around $T_{\rm eff}=1300$ K,
where it reaches about  10\% (again, at $K_{zz}=10^4$ cm$^2$ s$^{-1}$).
The lower panel displays the dependence of the non-equilibrium ratio
on surface gravity for two sets of models. In all models, $K_{zz}=10^4$
cm$^2$ s$^{-1}$. The upper set of models (diamonds connected by dashed
lines) are for the ``fast2'' CO/CH$_4$ reaction, while the lower set
of models (stars connected by dotted lines) are for the ``slow''
CO/CH$_4$ reaction. The curves are labelled by the value of $\log_{10} g$  
in cm s$^{-2}$. As discussed above, the lower-gravity models exhibit
larger departures from equilibrium for temperature below 1300 K. However,
the trend is reversed for $T_{\rm eff} > 1300$ K, where the lower gravity
models are closer to equilibrium. 

\subsection{Beyond the T Dwarfs}
\label{beyond}

In this section, we briefly explore the effects on spectra of departures 
from chemical equilibrium for brown dwarfs cooler than the coolest known
T dwarfs (see also Burrows, Sudarsky, \& Lunine 2003).  To this end,
we extend the grid of models described in \S\ref{res_lt}, and
compute models for $T_{\rm eff} = 200,\, 300,\, 400,\, 500,$
and 600 K.  For the models below 500 K, we include the effect 
of a water cloud with a modal particle size of 100 $\mu$m
and a super-saturation parameter of 1.0\%.
The other basic parameters are the same as in \S \ref{res_lt}.
In Fig. \ref{fig16}, we display the mixing time and the reaction times
for three models with $T_{\rm eff} = 200,\, 400,$ and 600 K,
for $g=10^5$ cm s$^{-2}$, $K_{zz}=10^4$ cm$^2$ s$^{-1}$, and the
``fast2'' CO/CH$_4$ reaction. In all models, the intersection 
of the mixing time and the reaction times occurs in the convection zone.
Therefore, the actual value of the diffusion coefficient $K_{zz}$, which 
influences the mixing time only at temperatures and pressures lower
than those at the intersection, is inconsequential. 
(In other words, $K_{zz}$ influences the $T_{\rm mix}$ curve only on the
left of the intersection point, which is already determined by the local
value of the convective velocity.)
Note that the intersections move to lower temperature with decreasing $T_{\rm eff}$.
The somewhat wiggly character of the $t_{\rm mix}$ curve in convection zone
follows form the fact the the convective velocity is given by (e.g., Mihalas 1978)
\begin{equation}
v = (gQH/8)^{1/2} (\nabla-\nabla_E)(l/H)^2\, ,
\end{equation}
where $g$ is the surface gravity, $H$ the pressure scale height,
$Q=(\partial\ln\rho/\partial\ln T)_P$, $l$ is the mixing length, 
and $(\nabla-\nabla_E)$ is a difference between the true temperature
gradient, $\nabla=\partial\ln T/\partial\ln P$, and the gradient
of the convective elements. This is computed by considering the
efficiency of convective transport. In the standard mixing-length
approach, the efficiencies are computed differently for Rosseland optical
depth smaller and larger than unity. The wiggly behavior corresponds to 
a transition between Rosseland optical depth larger than and smaller than 1.
In any case, the uncertainty in the exact value of the convective
velocity does not influence the non-equilibrium abundances of CO, CH$_4$,
and NH$_3$ in a significant way.

Figure \ref{fig17} depicts the abundance profiles for the models displayed
in Fig. \ref{fig16}. The non-equilibrium number fraction of CO sharply decreases mixing time
with decreasing $T_{\rm eff}$.  This is a consequence of the fact that
while the intersection points of the $t_{\rm mix}$ and the
$t_{\rm CO}$ curves occur at similar temperatures around 1100 K,
these intersection points are reached at much higher pressures for lower $T_{\rm eff}$.
Consequently, the equilibrium number fraction of CO at the intersection point
is lower. For ammonia, the equilibrium abundance profiles become flatter for  
deceasing $T_{\rm eff}$ and, therefore, the differences between the
equilibrium and non-equilibrium number fractions are smaller.

In Fig. \ref{fig18}, we depict the predicted flux for the three models
displayed in Figs. \ref{fig16} and \ref{fig17}. The hottest model, at
$T_{\rm eff}=600$ K, exhibits the same features as late T dwarfs (see \S\ref{res_lt}),
namely the non-equilibrium flux is higher in the 8$-$14 $\mu$m region (by about
15-20\%) and smaller in the the 4$-$5 $\mu$m region (on average, by about 10\%).
In the cooler model at $T_{\rm eff}=400$ K, the departures in the 8$-$14 $\mu$m region
due to non-equilibrium chemistry of ammonia are slightly smaller than in the 600-K model.
However, the differences in the 4$-$5 $\mu$m region essentially disappear because
the non-equilibrium mixing ratio of CO, though much higher than the
equilibrium one, is already quite small.  Therefore, CO contributes but 
little to the opacity in this region. Finally, for the coolest model 
at $T_{\rm eff}=200$ K, the non-equilibrium effects almost disappear 
in both the 4$-$5 $\mu$m and 8$-$14 $\mu$m regions.

To diagnose the trends described above, we display in Fig. \ref{fig19} the 
temperature/pressure profiles for the same models (both 
non-equilibrium and equilibrium) depicted in Figs. \ref{fig16} 
- \ref{fig18}. In Fig. \ref{fig20}, we display the ``radiation formation
temperature'' as a function of wavelength for the two extreme models with 
$T_{\rm eff} = 200$ and 600 K. The radiation formation temperature is defined as a local
temperature in the atmosphere where the monochromatic optical depth is 2/3.
This quantity indicates the position/layer from which a photon of a given wavelength
effectively decouples from the atmosphere.

\section{Model Comparison with Gliese 570D Observations}
\label{gl570d}

In Fig. \ref{fig21}, we display the observed spectrum of Gliese 570D, 
kindly supplied to us in digital form by M. Cushing (Cushing et al. 2006). 
This is one of the latest known T dwarfs, and is the object studied by Saumon et al.
(2006). We also display two theoretical models computed for  
$T_{\rm eff}=800$ K and $g=10^{5}$ cm s$^{-2}$, namely an LCE and a non-LCE model.
The non-equilibrium model is computed with $K_{zz}=10^4$ cm$^2$ s$^{-1}$,
and for the ``fast2'' CO/CH$_4$ reaction rate, but, as demonstrated in Fig. \ref{fig13},
this predicted flux essentially coincides with the flux computed for all other
values of $K_{zz}$ and CO/CH$_4$ reaction time prescriptions. The agreement between
the observed and predicted spectrum for non-LCE model is excellent in the
region between 8 and 14 $\mu$m, while the fit is much worse for the LCE model.
We thus confirm the results of Saumon et al.\footnote{Note that the predicted 
fluxes for both our models and those of Saumon et al. (2006) are too 
low between 7 and 8 $\mu$m, for both LCE and
non-LCE models. The reason for this is unclear, but may be related to
deficiencies in the extant water opacity databases.}.
In a future work, we will show comparisons between our non-equilibrium
spectral models and observations in the near- and mid-IR of other late T dwarfs.

\section{Discussion and Conclusions}
\label{conclusions}

In this paper, we have performed a systematic study of the spectral consequences
of departures from chemical equilibrium in carbon and nitrogen chemistry
in the atmospheres of L dwarfs, T dwarfs, and substellar-mass 
objects down to 200 K. Unlike previous investigations, the
temperature/pressure profiles of the non-equilibrium models are fully consistent
with the non-equilibrium chemistry.  Such models reveal that failure to 
calculate the non-equilibrium atmospheric profiles in a self-consistent fashion 
can result in errors of as much as $\sim$50$-$100 K in the atmospheric temperatures
at a given pressure.  Our effort has generated an extensive grid of  
non-equilibrium model spectra for effective temperatures
from 700 K to 1800 K (in steps of 100 K), for three values of surface gravity,
$g = 10^{4.5},\, 10^{5.0}$, and $10^{5.5}$ cm s$^{-2}$, for four values of the 
coefficient of eddy diffusion in the radiative zone
$K_{zz}= 10^2, 10^4, 10^6$, and $10^8$ cm$^2$ s$^{-1}$, and for three different
CO/CH$_4$ reaction time prescriptions. We have also provided 
clear and cloudy model variants.  We find, in keeping with previous studies, 
that there are essentially only two regions where the effects of departures 
from chemical equilibrium significantly influence the
predicted spectrum.  These are in the M ($\sim$4$-$5 $\mu$m) and N (8$-$14 $\mu$m)
bands due to CO and NH$_3$, respectively.  The overabundance of CO can be very large,
but translates into flux suppressions in the M band of at most $\sim$40\% between 
effective temperatures of 800 and 1800 K.  The effect is largest around
$T_{\rm eff} \approx 1100$ K.  The underabundance of ammonia can be 
significant, but translates into flux enhancements in the N band of no more than
$\sim$20\% for the $T_{\rm eff}$ range from 700 to 1800 K, with the largest
effects at the lowest values of $T_{\rm eff}$.  The magnitude
of the departure from chemical equilibrium increases with decreasing gravity,
with increasing eddy diffusion coefficient, and with decreasing speed of
the CO/CH$_4$ reaction.  Though these effects are modest, they lead to
better fits with the measured T dwarf spectra (see also Saumon et al. 2006,2007).

In addition, we have calculated an exploratory sequence of non-equilibrium models 
down to 200 K, including the possible effects of water cloud formation. 
These models show that in the region ``beyond the T dwarfs" (Burrows, 
Sudarsky, \& Lunine 2003) the suppression effect of non-equilibrium CO 
in the M band diminishes, then disappears, with decreasing $T_{\rm eff}$.

Hence, despite the modest suppression in the M band due to 
non-equilibrium CO chemistry for $T_{\rm eff}$s above $\sim$500 K,  
non-equilibrium effects do not compromise the use of the M band flux 
as a useful diagnostic of cool atmospheres and as one of their 
most prominent features.  At the higher $T_{\rm eff}$s above $\sim$500 K, this is 
due in part to the saturation of the augmented CO absorption, but importantly 
it is due to the fact that only about half of the M band overlaps with the 4.7-$\mu$m CO 
absorption feature.  At lower $T_{\rm eff}$s below $\sim$400 K, non-equilibrium effects in the  
M and N bands all but disappear. Overall, the prominence and importance of 
the M band flux for searches for the coolest dwarfs and for brown 
dwarfs (``Y" dwarfs ?) beyond the T dwarf range is not substantially affected
by non-equilibrium chemistry.

\acknowledgments

We thank M. Cushing, J. Budaj, and W. Hubbard for fruitful discussions 
and help during the course of this work. We also thank the referee, 
K. Lodders, for her detailed comments. This study was supported in part 
by NASA grants NAG5-10760, NNG05GG05G, NNG04GL22G and NNX07AG80G,
and through the NASA Astrobiology Institute under Cooperative
Agreement No. CAN-02-OSS-02 issued through the Office of Space
Science.

\newpage

\clearpage 

\clearpage

\clearpage

\clearpage 

\clearpage

\clearpage

\clearpage

\clearpage

\clearpage

\clearpage

\clearpage

\clearpage

\clearpage

\begin{figure}
\plotone{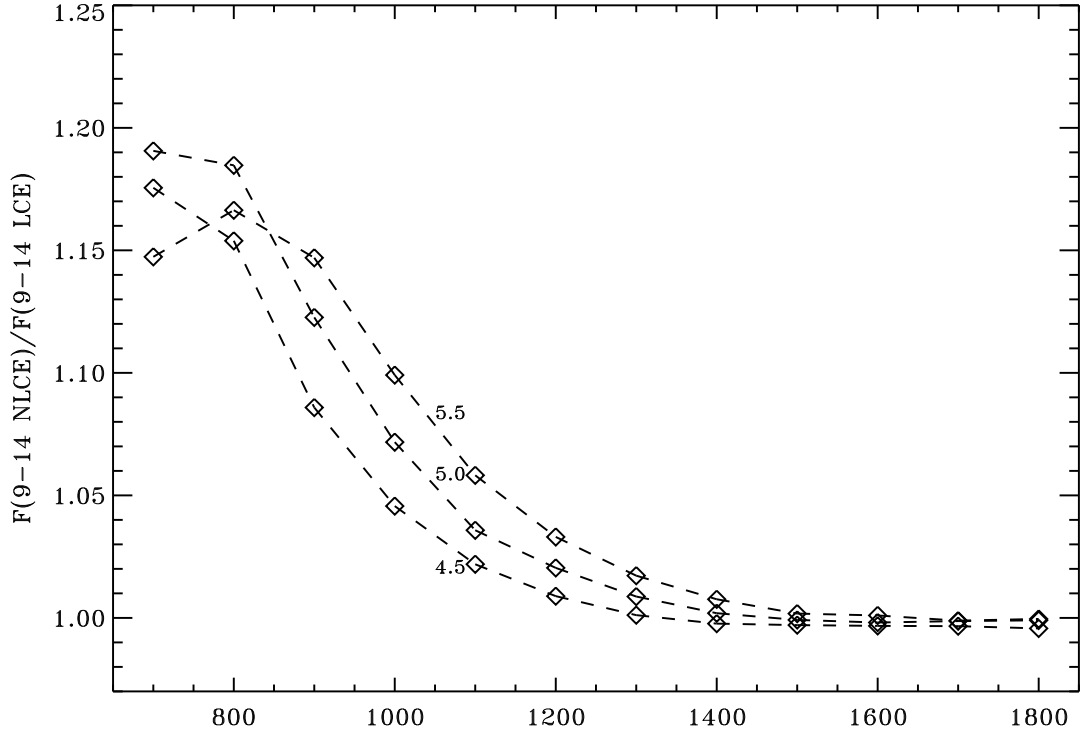}
\caption{Similar to Fig. \ref{fig14}, but for the ammonia region between
9 $\mu$m and 14 $\mu$m. Since the predicted spectra are insensitive to the
speed of the CO/CH$_4$ reaction, and only weakly sensitive to $K_{zz}$,
we display only models with $K_{zz}=10^4$, and the ``fast2'' CO/CH$_4$
reaction. The curves are labelled by the value of $\log_{10} g$ in
cm s$^{-2}$.
}
\label{fig15}
\end{figure}

\clearpage

\begin{figure}
\plotone{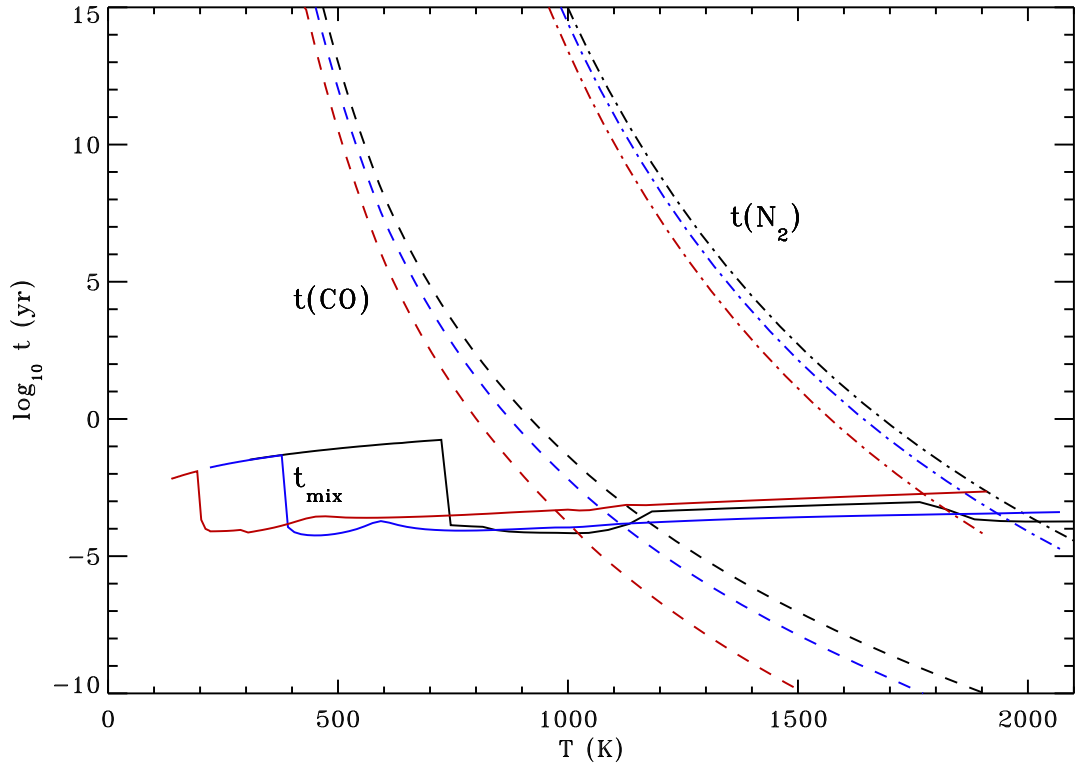}
\caption{A plot of the mixing time (solid lines), the reaction time
for the CO/CH$_4$ (dashed lines), and the reaction time for
the NH$_3$/N$_2$ reaction (dash-dotted lines), for models with
$g=10^5$ cm s$^{-2}$
$K_{zz}=10^4$ cm$^2$ s$^{-1}$, and $T_{\rm eff} = 600$ K (black lines),
400 K (blue lines), and 200 K (red lines).
}
\label{fig16}
\end{figure}

\clearpage

\begin{figure}
\plotone{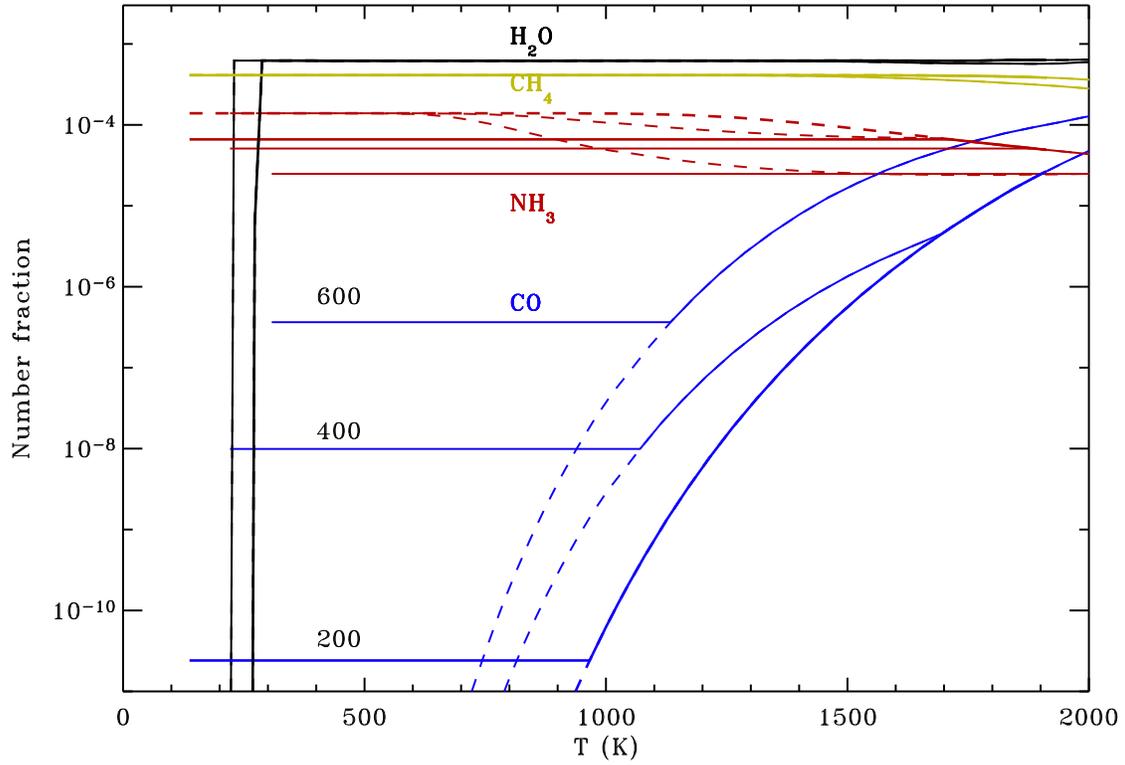}
\caption{A comparison of the equilibrium (dashed lines) and non-equilibrium
(solid lines) abundances of water (black), methane (yellow),
ammonia (red), and CO (blue), for the same models as displayed in
Fig. \ref{fig16}. Thin lines depict models for $T_{\rm eff} = 600$ K,
thicker lines for for $T_{\rm eff} = 400$ K, and the thickest lines for  
$T_{\rm eff} = 200$ K. The curves for CO are labeled by the value of $T_{\rm eff}$
in K.
}
\label{fig17}
\end{figure}

\clearpage

\begin{figure}
\plotone{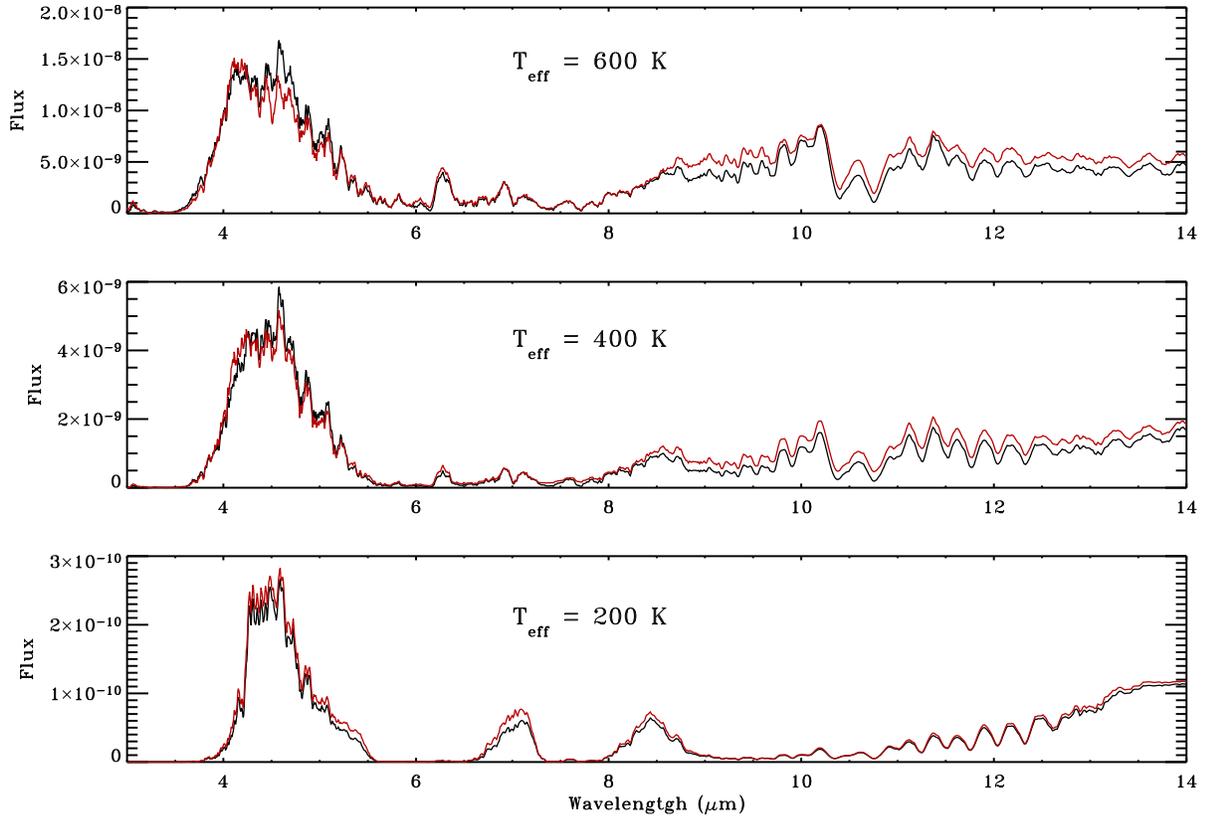}
\caption{Predicted flux for models with equilibrium (black) and
non-equilibrium (red) models with $g=10^5$ cm s$^{-2}$
$K_{zz}=10^4$ cm$^2$ s$^{-1}$, and $T_{\rm eff} = 600$ K (upper panel),
400 K (middle panel), and 200 K (lower panel).
}
\label{fig18}
\end{figure}

\clearpage

\begin{figure}
\plotone{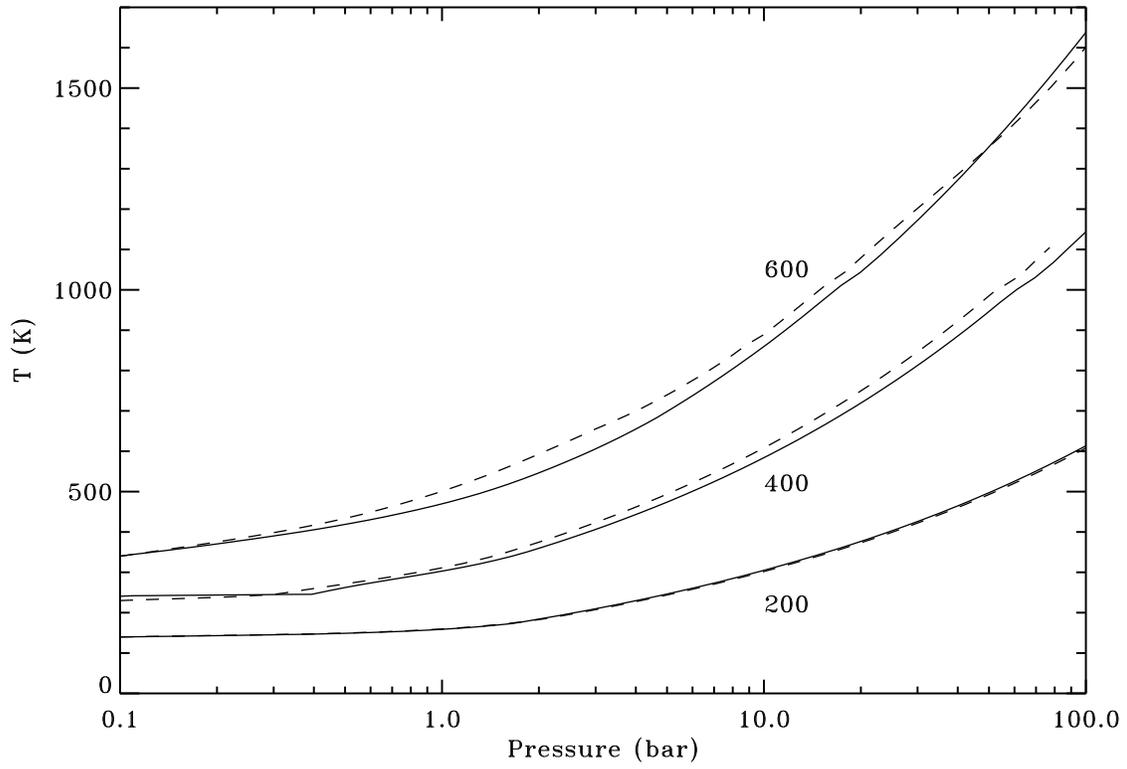}
\caption{A comparison of temperature-pressure profiles for equilibrium (dashed lines)
and non-equilibrium (solid lines) models of $T_{\rm eff}=600,\, 400$,
and 200 K (from top to bottom). $g$ is equal to $10^{5}$ cm s$^{-2}$.
}
\label{fig19}
\end{figure}

\clearpage

\begin{figure}
\plotone{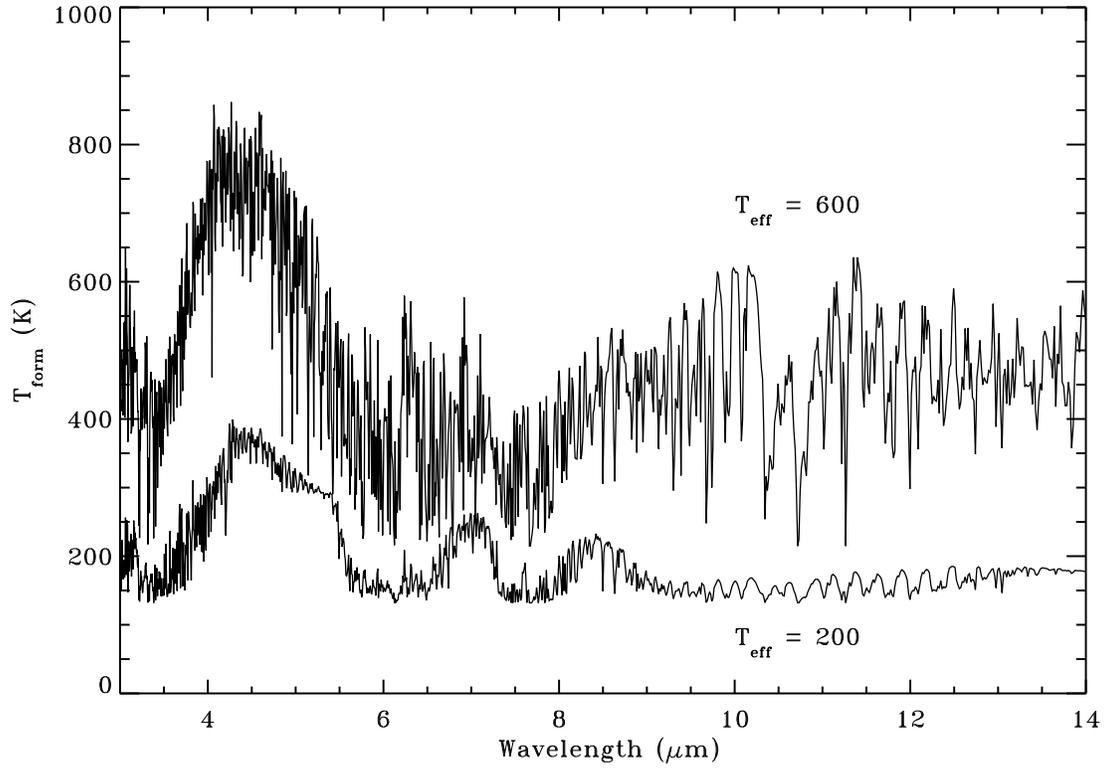}
\caption{Radiation formation temperature as a function of wavelength for models
with $T_{\rm eff}=600$ K (upper curve) and 200 K (lower curve).
$g$ is equal to $10^{5}$ cm s$^{-2}$.
}
\label{fig20}
\end{figure}

\clearpage

\begin{figure}
\plotone{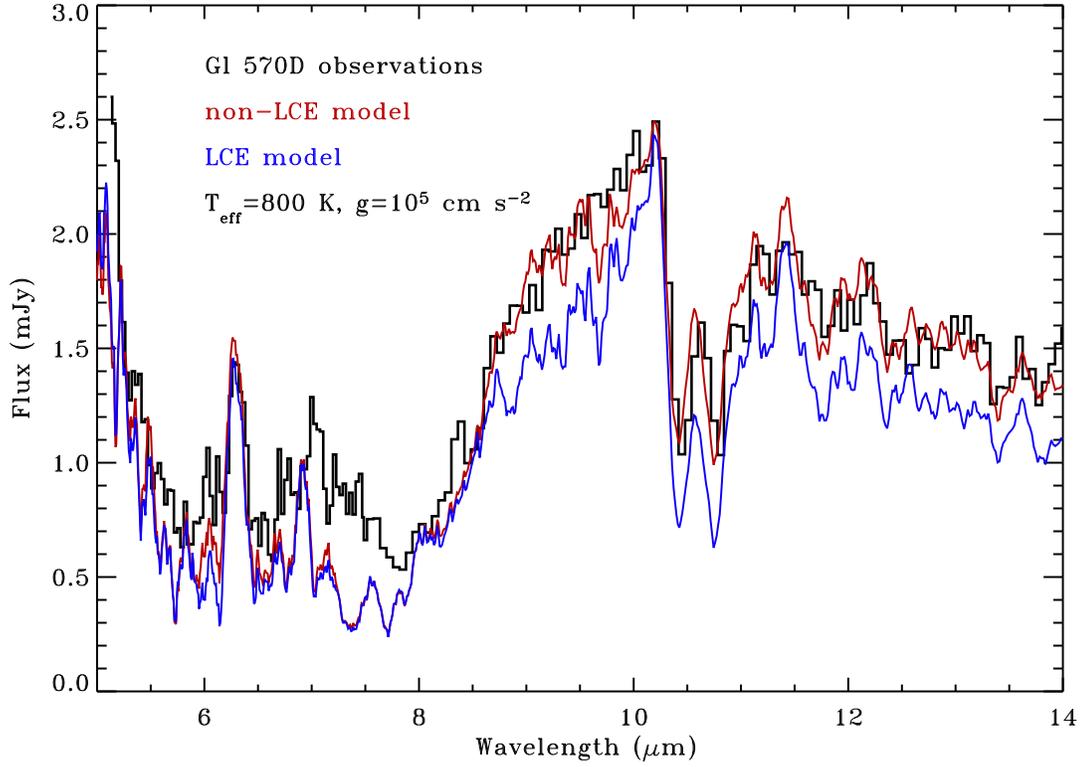}
\caption{Comparison of observed and predicted spectra of Gl 570D.
Thick black histogram displays the observed flux, the red and blue
lines are the predicted spectra for non-equilibrium (red) and equilibrium
(blue) models, computed for $T_{\rm eff}=800$ K and $g=10^{5}$ cm s$^{-2}$.
The non-equilibrium model displayed is computed with $K_{zz}=10^4$ cm$^2$ s$^{-1}$,
and for fast CO/CH$_4$ reaction rate, but, as demonstrated in Fig. \ref{fig13},
this predicted flux essentially coincides with the flux computed for all other
values of $K_{zz}$ and CO/CH$_4$ reaction time prescriptions.
}
\label{fig21}
\end{figure}

\end{document}